# End-to-end science operations in the era of extremely large telescopes

**Olivier R. Hainaut,[1,a] Marie Lemoine-Busserolle,[2,b] Christophe Dumas,[3,c] Robert W. Goodrich[4,d], Bryan W. Miller,[b] Michael F. Sterzik,[a] Thomas Bierwirth,[a] Sidney Wolff,[b] Andrew W. Stephens,[b] Gelys Trancho,[c] Warren Skidmore,[c] Kim Gillies[c]**

[a] ESO, Karl-Schwarzschild-Straße 2, 85748 Garching-bei-München, Germany.
[b] NSF's NOIRLab, 950 N. Cherry Ave, Tucson, AZ 85719, USA.
[c] TMT International Observatory, 100 W. Walnut St., Suite 300, CA 91124, USA.
[d] GMTO Corporation, 465 N. Halstead St., Suite 250, Pasadena, CA 91107, USA.

**Abstract:** Observatory end-to-end science operations is the overall process starting with a scientific question, represented by a proposal requesting observing time, and ending with the analysis of observation data addressing that question, and including all the intermediate steps needed to plan, schedule, obtain, and process these observations. Increasingly complex observing facilities demand a highly efficient science operations approach and at the same time be user friendly to the astronomical user community and enable the highest possible scientific return. Therefore, this process is supported by a collection of tools. In this paper, we describe the overall end-to-end process and its implementation for the three upcoming extremely large telescopes (ELTs), ESO's ELT, the Thirty Meter Telescope (TMT), and the Giant Magellan Telescope (GMT).

## 1. End-to-end operations

### a. History

Historically, astronomers obtained observing time on their local or national telescope, travelled there to perform the observations (sometimes involving an arduous trip and an extended stay), and returned to their institute with stacks of photographic plates. Over time, the main change to this paradigm was to replace plates with electronic detectors, recording their output on magnetic tapes or optical discs. Each observer would acquire the ancillary data required to calibrate their observations and, while the plates (or tapes) formally remained the property of the observatory, little or no effort was made to make these available to the broader community. Some exceptions existed — for example, the extensive Harvard plate collection[1].

For ground-based astronomy, this changed with the 8- to 10-meter telescopes built in the 1990s[2]. In addition to the traditional v*isitor mode* (VM, also known as *classical observing*) described above, ESO, Gemini, and others introduced s*ervice mode* (SM, also known as *queue mode*): observations are defined in advance, then performed by observatory staff optimizing the schedule of the telescope and matching the observing conditions with the requirement of each observation.

[1] ohainaut@eso.org
[2] marie.busserolle@noirlab.edu
[3] cdumas@tmt.org
[4] bgoodrich@gmto.org

In parallel, the observatory started to calibrate the instruments systematically and consistently (setting up detailed calibration plans for all the instruments), and to methodically archive all data produced.

The scientific process of observational astronomy is cyclical, and it was realized that the new observing modes would only be successful if observing program information were managed from start to end. Quinn[3] and Puxley et al.[4] described a similar process which starts with a scientific question addressed by a request for observing time, proceeds with the observations prepared, executed, processed, analyzed, and finally, the scientific questions (hopefully) answered with a publication, which will trigger new questions. This end-to-end process enables efficient science operations of complex instrumentation, and at the same time enables an optimal scientific return for the community. ESO's current implementation is illustrated in Figure 1. The concepts it describes are relevant to most observatories.

This paper reviews and compares how the end-to-end operations are foreseen for the next generation of Extremely Large Telescopes (ELTs).

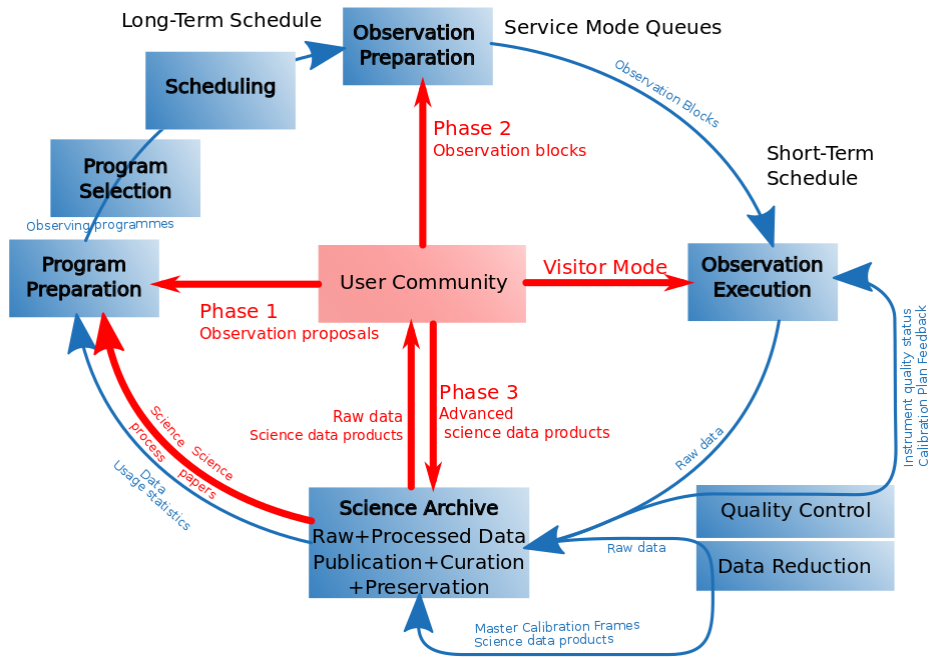

Figure 1. End-to-end operation process, as implemented at ESO (and very similarly for the other ELTs). The blue boxes represent the main sub-processes; the arrows are labelled with the information flowing through them. The red arrows show the main interactions with the community. Evolved from Quinn (1996).

## b. Evolution

At ESO, the end-to-end process was formalized for the VLT in the 1990s as the original Data Flow System (DFS) described by Quinn[3] and deployed on the NTT in 1996. It was subsequently upgraded to become an operational prototype of the VLT. The DFS was — and still is — built around the concept of the Observation Block (OB), which is the atom holding all the information required to perform an independent set of observations on a target. An OB consists of a series of *templates*, each defining a type of observation (e.g., to acquire a set of adaptive optics (AO) images with offsets, or to acquire long-slit spectra on and off-target). The OB only lists the *template parameters* (e.g., exposure times, filter, grating and wavelength, size and number of offsets, AO configuration...); the actual sequence of instruction to be performed using these parameters exists only in the instrument. While the OB is the atomic unit defining the minimum dataset making scientific sense, the template is the indivisible quantum in terms of the calibration plan (as an example, the OB includes templates describing a series of images in R and V, for a program aiming at measuring colors; for the calibration plan, the R and the V filters must both be calibrated). The templates, the definition of their parameters and their allowed ranges (together with a series of additional configurations files, defining for instance the relevant atmospheric constraints), constitute the Instrument Package (IP), which fully describes the characteristics and capabilities of the instrument.

Originally, the various subprocesses shown in Figure 1 were implemented with relatively independent tools. Over 25 years of operation, the DFS evolved, integrating new concepts (e.g., *scheduling containers* grouping OBs, for instance for large surveys on the VST and VISTA), new observing modes (e.g., *Rapid Response Mode* for targets of opportunity requiring extremely urgent observations), new instruments with special requirements (e.g., ESPRESSO, which can be used during the same night on any of the four VLT units, or use all four simultaneously), and the production of science-ready processed data. The tools grew organically, and some of the underlying technologies aged or even became obsolete. Nevertheless, the DFS still provides a common interface to around 25 instruments mounted on the 4 VLTs and several other smaller telescopes. This homogeneous system in science operations is the counterpart to the control system, homogeneous over all systems in terms of hardware (via the *VLT Standard*) and software (thanks to the standard *Central Control Software*).

In preparation for the ELT, ESO reviewed in depth the concepts at the core of the DFS, considering the new requirements from the ELT and the stipulation that the ELT must be fully integrated with the VLT operations. Requirements from new instruments (e.g., adaptive optics and laser guide stars were at their infancy at the beginning of the VLT, they are now ubiquitous), technology obsolescence (and the opportunities provided by recent technologies), and — not least — requests from the community were also considered. Highlights of this review are summarized in a series of Messenger articles ([5] for SM; [6] for the Archive; [7] for Phase 3; [8] and [9] for scientific return). The result of the review was that the fundamental concepts of the DFS (in particular the IPs, OBs and templates) were sound and would meet all the new requirements. While originally used only as an interface between the instrument control system and the observation handling system at the observatory, the use of IP is now being generalized across the entire DFS as the single, authoritative "hardware abstraction layer." It represents the physical properties of the instrument and reflects its actual (and eventually changing) capabilities, ensuring a consistency over the entire system that lacked in the original implementation. This gives ESO the opportunity to modernize the tools implementing the DFS, integrating them better and building them so they can support operations over an expected lifetime of at least 20 years. A coherent development approach was considered with the onset of the "DMO/DFS program" in 2013 to coordinate these efforts; a high-level, if slightly outdated, description of this plan was published by Hainaut et al. [10].

The United States Extremely Large Telescope Program (US-ELTP) has been formed as a collaboration of the Thirty Meter Telescope International Observatory (TIO), the Giant Magellan Telescope project (GMTO Corporation), and the NSF's National Optical-Infrared Astronomy Research Laboratory (NOIRLab). US-ELTP's main goal is to provide the user services to take full advantage of the capabilities of GMT and TMT by providing training, assistance, and integrated user-facing software tools. NSF's NOIRLab, together with TMT and GMT, developed a detailed plan for the user support system over all phases of the end-to-end process –called Scientific Data Life Cycle (SDLC) in this context, and visualized in Figure 2— and have allocated the responsibility for each action that must take place.

The Gemini Observatory is now a part of NSF's NOIRLab. Its expertise and the evolution of its operations therefore naturally served as a precursor on which the implementation of the SDLC tools builds extensively. The evolution of the Gemini science operations followed a similar trajectory to that of ESO's. All the original tools have evolved substantially, for example the Phase I Tool (PIT) and International Time Allocation Committee (ITAC) software were rewritten, the archive was replaced with the Gemini Observatory Archive running on Amazon Web Service[11], and the data reduction software is being transitioned from the Gemini IRAF package to the DRAGONS Python pipeline environment[12]. Gemini's original operations model envisioned 50% visitor/classical mode and 50% service/queue mode. However, investigators quickly voted for queue and now over 90% of the observing is done using some form of service mode. Gemini then implemented new observing and proposal modes —rapid target-of-opportunity (ToO) mode with programmatic access, fast-turnaround, and large-long programs— in order to enhance flexibility for programs of different sizes and to support student training[13]. Also, after years of usability improvements to the Observing Tool the point was reached where new required capabilities were too difficult to implement in the existing infrastructure. Therefore, NSF's NOIRLab/Gemini has initiated the Observatory Control System (OCS) Upgrades Program to reimagine and replace most of the high-level operations applications with modern, web-based tools[14]. The Gemini Program Platform (GPP) is the core of this system and will unify proposal preparation, integration time calculators, and observation definition in a single web app called Explore. A single, cloud-based, observing database will unify Gemini North and South and allow them to work together via an automatic, real-time scheduler.

Building off the Gemini experience and work on GPP, NSF's NOIRLab will build the US ELT Program Platform (UPP), which will provide a significant subset of the functionality within the overall SDLC. The SDLC also includes, of course, the execution of observations and data collection by GMT and TMT. The UPP will provide a common user interface that will support applications for observing time on either or both the TMT and GMT and will make it possible to compare the performance of the imagers and spectrographs on each in order to determine the best strategy for specific programs. The development of these user services will be carried out by NSF's

NOIRLab and TMT and GMT partners, as well as the US community. Similarities to NSF's NOIRLab/Gemini include access to two telescopes, one in each hemisphere; demanding observations that require special atmospheric conditions for adaptive optics and thermal infrared observations; a complex suite of instruments, several of which are available on any given night; and extensive use of queue and remote observing. Therefore, in the sections that follow, we will reference NSF's NOIRLab/Gemini current capabilities and planned improvements when we describe the end-to-end process for GMT and TMT.

In Section 2, we will go through the various sub-processes of the end-to-end science operations, and compare the plans for ELT, GMT and TMT.

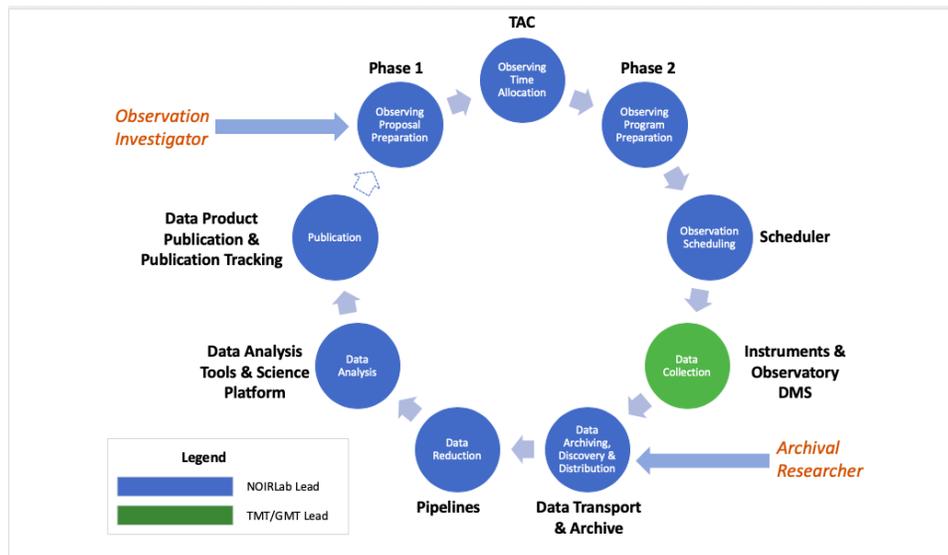

Figure 2. The Scientific Data Life Cycle (SDLC) and the main phases that the UPP will support.

## c. *Technologies*

Over the past 20 years, tools supporting end-to-end operations have been implemented as software that users would install on their own computers. The wide zoo of operating systems used in the community resulted in complications for the developers (and often frustration for the users) and presented obstacles to the deployment of updates and bug fixes. New technologies, however, support platform-independent software access, via server-side software, or exportable self-contained "containers," or other means.

ESO's strategic choice was to move towards online and web-based user interfaces, that allow efficient and platform-independent development. These interfaces are connected via APIs to servers hosted by the organization. Scripts can also use the APIs, allowing power users to automate workflows requiring long or repetitive interactions with the web interfaces, and even to develop custom user interfaces. Having the "business logic" running on ESO servers has the advantage that updates and bug fixes are deployed on the server and do not require the user to download or install new software. Also, with a strict separation between user interface and the business logic, the interface can be updated or even completely overhauled without changing the server-side services. The technology powering web user interfaces evolves more rapidly than the core software technology.

NSF's NOIRLab's UPP will also include various web-based applications, web-based user interfaces and a collection of secure APIs to all services to facilitate advanced user tasks, reporting, and automation. Some of these applications will be externally available to users while others will be for NSF's NOIRLab, GMT and TMT internal use only. It is a requirement that most of the user-facing applications will be used for both GMT and TMT, therefore a cloud-based deployment model has been selected.

*d.  User Portal Infrastructure*

Many, if not most, of the interactions between the users and the services provided by the observatory benefit from the users identifying themselves, to have access to their personal information such as the proposals they submitted, or the data they collected.

At ESO, this service is provided by the User Portal[15], which holds a profile for each user, including their basic identity, their professional status, institution affiliation, year of PhD, and gender, as well as a series of keywords describing their scientific and technical expertise. The user's profile also includes the list of specific roles and privileges they hold (for instance, referee for the time allocation committee [TAC], or operation staff astronomer), which are used to control access to the various confidential material in the system. Keywords capture the scientific and technical expertise of each individual (used, for example, to identify suitable referees for the TAC), and their affiliation to flag conflicts of interest. Affiliation, seniority, and gender are used to quantify and monitor the diversity of the community and of the TAC. The whole User Portal system was implemented following the EU General Data Protection Regulation. Using their User Portal identification, the user can log on all the ESO services, and the User Portal itself acts as a gateway to these services, as illustrated in Figure 3.

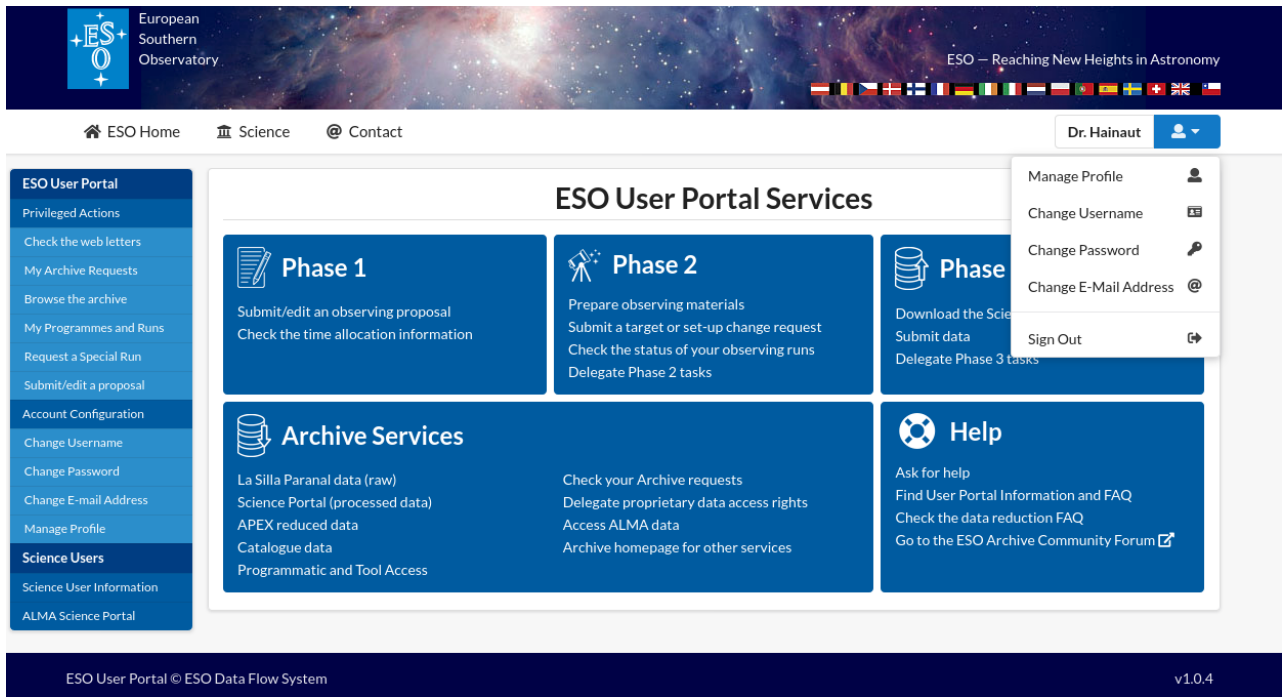

Figure 3. The ESO User Portal landing page, from which a user has access to most services provided by ESO.

NSF's NOIRLab's UPP will provide publicly available information to unauthenticated users but logging in is required to submit proposals, for time allocation process, to prepare observations, and to access proprietary information. The UPP will provide a mechanism to manage user roles and permissions.

Similarly, the UPP will provide a common web interface to the user-facing applications. In addition, the UPP Common Software Services are focused on providing a high-quality experience for users. There will be a user-facing dashboard along with administrative tools to configure user accounts, define collaboration groups, and authenticate data-access permissions. This system will also notify users of the status of their proposals and of approved observations that are in progress. Users will access all UPP services using a common single sign-on (SSO) system. There will be a Help Desk system that may be used by the observatories and their partners. Over time, the Help Desk team will develop a knowledge base to facilitate support. Metrics will track performance, including program completion rates, archival data use, and publications.

## 2. End-to-end operation sub-processes

In the following sections we will discuss and compare the various tools and steps of the end-to-end operation process, and how they will be implemented for the ELTs.

### a. Exposure Time Calculators

A preliminary stage to any project consists in selecting the instrument (and the configuration of that instrument) that will collect data suitable to address the scientific question at hand. While experienced observers get a feeling of what magnitude is too faint to be observed, complex instruments and/or novice users require a more quantitative approach. With the *exposure time* *calculator* (ETC), the astronomer can compute the exposure time required to reach the desired signal-to-noise ratio (SNR) on their targets. The ETCs can simulate a range of objects (stars, galaxies, quasars...), observing conditions (seeing, moon, airmass), and sky transmission and brightness. They include transmission curves for all the optical elements involved and can account for various point-spread functions (or line-spread functions for spectrographs). While the ETC is meant to provide a precise estimate of the SNR for a target, more complex *instrument simulators* can produce realistic data frames as they would be produced by an instrument, including noise features, point- or line-spread-function, etc. There is a continuum between the simplest ETC and the most complex simulator.

At ESO, the ETCs have been web-based since their early days. They are being completely overhauled to the new technology standards, and their calculation engine is upgraded with the goal for the results to reach a 10% accuracy. The new ETC web interface[16] is designed so that users unfamiliar with the instrument do not need to read the full documentation to determine whether the instrument is suitable for their needs (see Figure 4 for an illustration). Thanks to the APIs and to the fact that the ETC uses the same IP as the other tools, the Phase 1 and Phase 2 preparation tools (see below) will be able to call it directly to refine the SNR and exposure times. Also, the APIs allow users to programmatically scan parameters, or to go through a list of targets. Finally, the APIs make the simulations available to the quality control process (see below), for instance to compare the expected SNR with that of the observations.

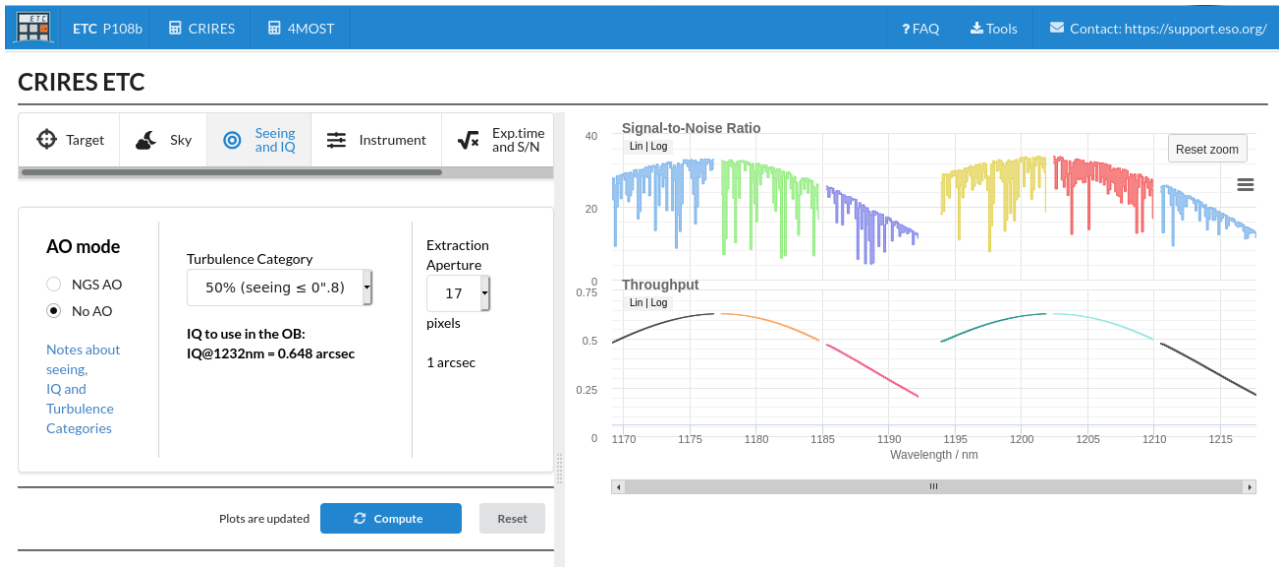

Figure 4. An example of Exposure Time Calculator: a zoom on a couple of spectral orders of ESO's CRIRES.

NSF's NOIRLab/US-ELTP is planning to build the system that will allow scientists to develop and write proposals for GMT/TMT observations based on the NSF's NOIRLab/Gemini "Explore". Potential investigators will enter only the target details and their science requirements (e.g mode, wavelengths, resolution) in the web-based tool, called *"Prepare",* which will be tightly integrated with the ETCs. They will not have to worry about which instruments they need or what GMT/TMT will have available in a specific semester. "*Prepare*" will automatically provide a list of configurations that will deliver the requirements that were requested, will display for each configuration how long it will take to reach the required S/N, and will present the ETC output for the selected configuration. This should help the investigators verify that they will get what they expect (see Figure 5).

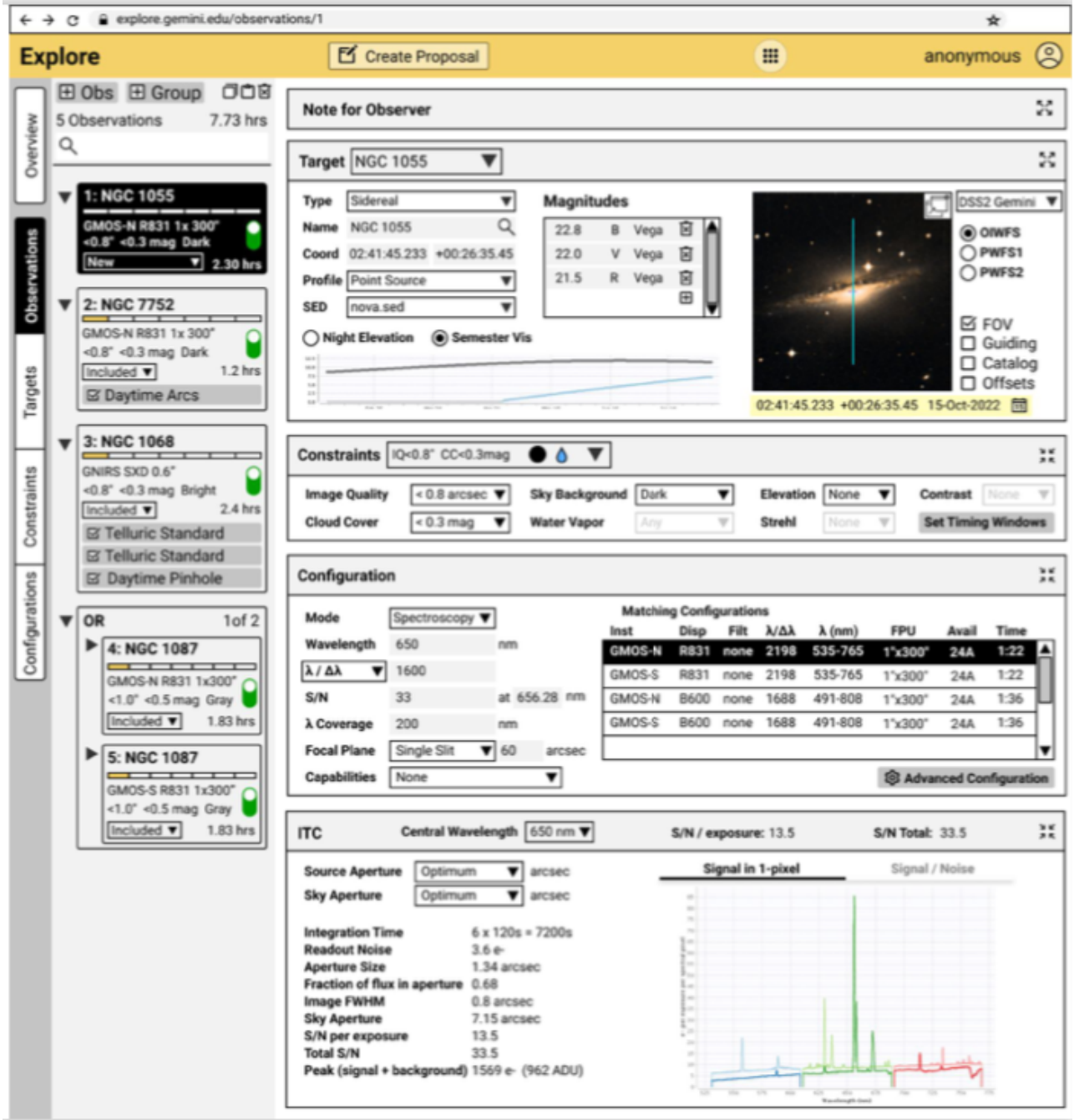

Figure 5. Mockup of the Observation view of the GPP/Explore application, the precursor of the UPP/Prepare, with a single observation highlighted and showing the list of Matching Configurations satisfying the high-level science requirements as well as the Instrument Time Calculator (ITC) information.

### b. Phase 1

At Phase 1, the astronomer submits a proposal to the observatory, making their science case as a narrative that will be evaluated by the scientific referees. The proposal also includes technical information on the observations they want to perform, the instrument set-up required, and the amount of time requested.

At ESO, the former system was based on a LaTeX template whose content was parsed to feed a database. The limitations imposed by LaTeX and the maintenance issues caused by the series of organically grown tools handling the proposals triggered a complete overhaul of the Phase 1 system. The new system — P1[17]— is based on a web interface that populates the database directly using APIs. This technology choice supports collaborative editing of a proposal by all co-investigators, identified via their User Portal profile. To assemble the technical information, the astronomer associates astronomical targets to instrument set-ups defined using templates from the Instrument Package (again, the same authoritative instrument model), and assigns a total exposure time and overhead. This results in simplified OBs that contain all the information required to evaluate and schedule the observations, but without the burden of fully detailing the observations. This also ensures that inconsistent configurations can simply not be defined. The system does the administrative work of keeping track of all the set-ups and total time. As all the information is directly entered into the Phase 1 database, a variety of reports are produced, including the traditional "proposal" document (as a PDF file) that comes in various flavors, including a fully anonymized version to be used by the referees. The web technology used in the interfaces enables various tools to be directly built into the system: visibility plots of the targets, consistency checks (see Figure 6). Ultimately, the Phase 1 interface will be directly connected to the ETC. This will allow the user to seamlessly compute the exposure time required from their proposal, and to store the results of the ETC in the proposal, simplifying the technical feasibility review by observatory staff. The Phase 1 system is designed to support all the proposal types envisioned: the traditional 6-month cycles currently used at ESO (possibly extended to 12-month in the future), a fast-response director discretionary time program, special calls for proposals for science verification of new instruments or to cover unexpected situations, and an upcoming fast turnaround scheme with a monthly cycle.

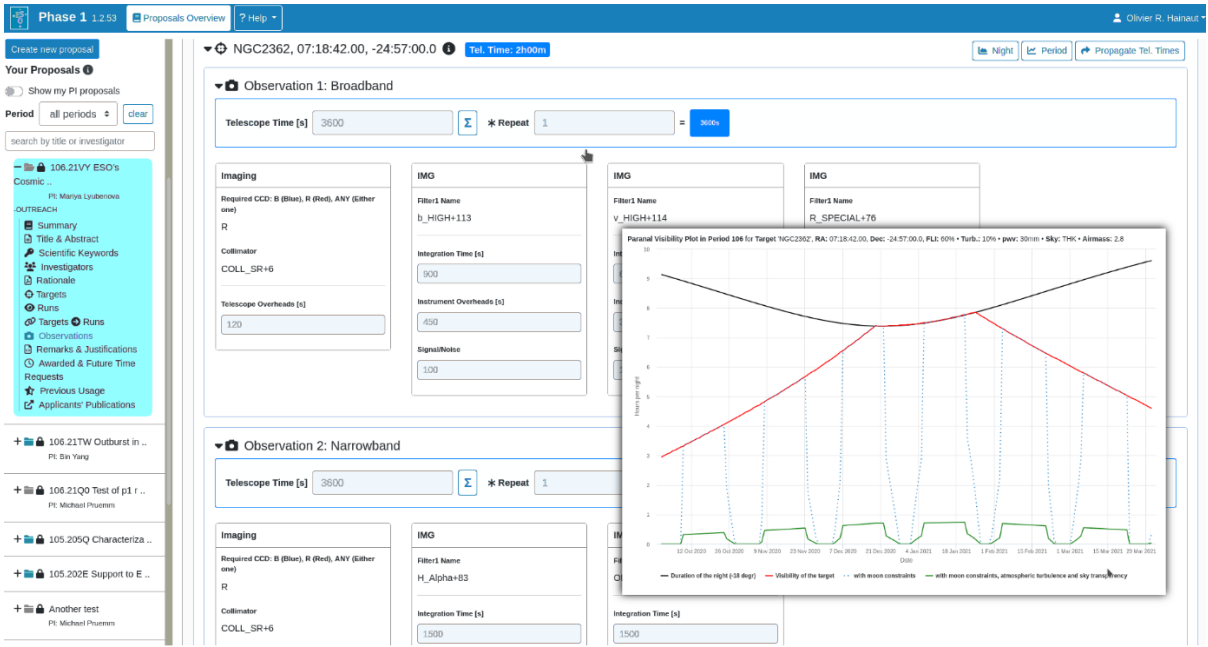

Figure 6. Screenshot of ESO's P1web interface, showing the definition of an observation. The inset displays the object visibility and probability of realization of the requested observing conditions over the semester.

In the NSF's NOIRLab/UPP system investigators will use the "*Prepare*" application as described above to complete their proposals. After users enter targets and science requirements and select the instrument configurations from the list of matching options, "Prepare" will calculate the total time needed for the project using the built-in ETC and knowledge of all overheads. The investigators will be able to work collaboratively on the proposal, with the principal investigator (PI) being able to manage permissions. They will enter the list of co-investigators and define the time requests from each of GMT's and TMT's participants. The scientific justification, experimental design, and other "essay" sections will be created using provided LaTeX or Word templates and the resulting PDF files will be uploaded and included with the proposal (see Figure 7). UPP will support dual-anonymous review processes (DARP) for reducing unconscious bias. The teams will be able to submit or retract proposals at any time up to the deadline. As with the ESO system, UPP/Prepare will support a variety of proposal types and multiple, simultaneous calls-for-proposals.

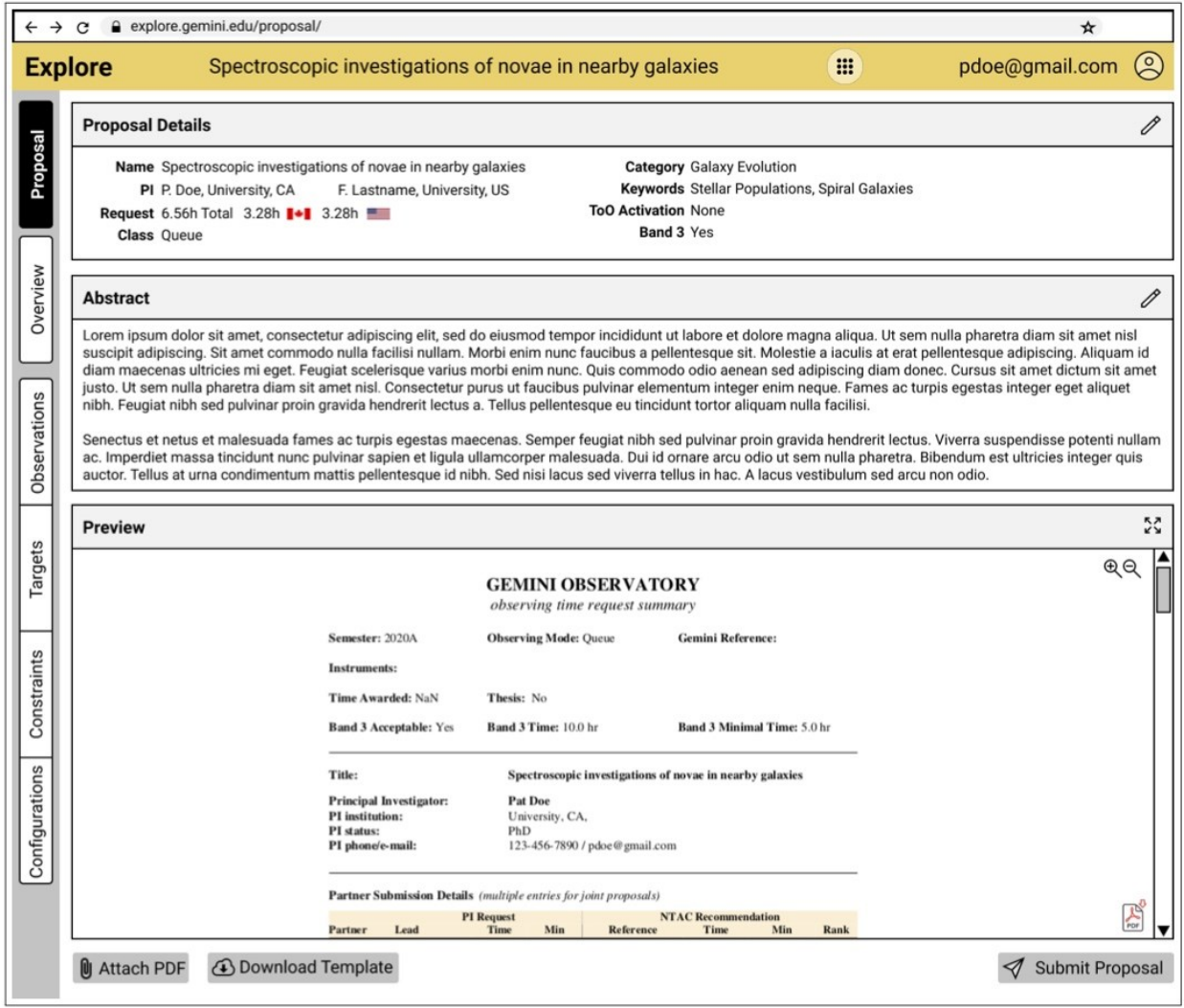

Figure 7. Mockup of the Proposal view of the GPP/Explore application, the precursor of the UPP/Prepare.

*e. Review, time allocation and (long-term) Scheduling*

The number of proposals submitted for large telescope results in a strong oversubscription. In the case of the VLTs, the oversubscription is typically in the range of 3 to 7, and we can expect values at least as high for the forthcoming ELTs. This implies that the proposals must be thoroughly reviewed, and telescope time can be granted only to the best ones.

At ESO, the evaluation process is separated in a scientific assessment performed by an Observing Program Committee (OPC, called Time Allocation Committee or TAC by other organizations) composed of astronomers from the community, and a technical evaluation by the observatory staff aimed at flagging proposals that are not feasible or whose time request is wrong by a large margin. A series of new administrative tools is being developed to assemble the OPC, to keep track of suitable referees, of their expertise, but also of their affiliation to flag conflicts of interest and ensure a balanced and diverse distribution of nationalities and genders. The tools will also provide the platform for the referees to access the proposals they need to review, and to deliver their grades and notes. These are then combined across the whole stack of proposals, resulting in a consolidated ranked list of proposals that will serve as the basis for time allocation. The administration tools are currently being developed, building upon P1, which collects the proposals, and the Users Portal, that centralize the personal information of proposers and referees.

At ESO, the time allocation itself is the result of scheduling the proposals starting from the highest ranked one and going down the list until all the available time is used up, taking into account the ESO Science Operation Policies[18] and the constraints imposed by the proposal themselves (time constraints, moon, seeing, ...). To provide additional flexibility, a moderate amount of oversubscription is allocated with a lower priority status, and programs with very loose observing constraints can also be granted time to serve as "fillers." The final schedule is reviewed by the ESO Director General who then formally allocates the telescope time. A new time allocation system is being developed, to address shortcomings of the current ones (which is slow and cumbersome, and deals with each telescope independently), and to support simultaneous or coordinated scheduling of more than one telescope. This new system is designed to schedule the simplified OBs prepared at Phase 1 together with the fully detailed OBs (prepared at Phase 2, below) that are already approved (which will be critical for scheduling fast turnaround proposal).

The current operations concepts for TMT and GMT call for each partner to operate its own TAC, with inputs from the various TACs merged into a single schedule by each observatory. All the proposals will be submitted into a common database maintained by NSF's NOIRLab, and the UPP will allow each partner to access the proposals submitted by its community of users. The partners will have the option of using the NSF's NOIRLab system to manage their TAC processes, devising their own independent process for review, or downloading the NSF's NOIRLab software and modifying it to meet their unique needs, as partners may develop their own criteria for evaluating proposals. NSF's NOIRLab's system will be designed to be flexible enough to encourage direct use of their software. NOIRLab will work with its community to develop criteria for assessing the merit of Key Science Programs and Discovery Programs, including scientific quality, data management, research inclusion, and other relevant factors. Different factors may apply to the evaluation of programs of different scales. In addition, the initial phase of review will be double-blind — proposers and reviewers will not be made known to each other.

Following their TAC process, each partner is required to provide to GMT and TMT a list of the proposals that it has approved for scheduling. This list will include the amount of time allocated and the proposal ranking. The ranked programs from multiple partner TACs must then be merged into a single integrated and prioritized observing list for each observatory. The merging TAC will include representatives from each individual TAC, to assure consensus on the result. It is important to note that the multi-partner proposal merging process for GMT and for TMT will likely be different. TMT is planning a multi-partner proposal merging process similar to the NSF's NOIRLab/Gemini International Time Allocation Committee[19], while GMT is thinking of a hybrid approach where each partner has the option to divide their share of GMT observing time between proposals to be selected by the GMT-wide ranking subpanels and proposals to be selected and ranked through their own internal process. NSF's NOIRLab will provide the full merging and scheduling tool for TMT and a visualization tool for GMT. The observatories themselves, using these tools, will be responsible for producing an integrated observing list and assigning long-term scheduling priority to the observations.

*f. Phase 2*

The principal investigators who receive the good news that their proposal is approved and scheduled must then prepare the observations in full detail. Whether the observations will be performed in VM or SM simply changes the time when this is done: well in advance for SM, while more time may be provided for VM observations.

At ESO, a single web tool, P2[20], is used for the preparation of the OBs for all the visible and IR instruments on all the telescopes. Thanks to the abstraction level provided by the IP, its software does not contain *any* instrument-specific code. In addition to the obvious advantages of maintainability and ease to add support of new instruments, this also gives the users a homogeneous way to prepare their observations across the whole fleet of over 20 instruments. With P2, the users prepare their OBs and store them directly in the database at the Garching headquarters. A bi-directional database replication ensures that the OBs are always the same in the Garching database and those on the La Silla and Paranal observatories. Furthermore, as P2 is built on documented public APIs[21], the OBs can be prepared programmatically, either using Python scripts or custom-built user interfaces tailored to the specific needs of the observing program. Nevertheless, some complex instruments require additional preparation, whose complexity cannot be described by the general structure of the IP templates. This preparation is done using an additional tool, ObsPrep, whose interface is integrated within P2 (see Figure 8). A secondary level of abstraction, which covers concepts common to several instruments (e.g., to define a complex sequence of offsets, or to select a reference star) is available to all instruments in ObsPrep. Complex functions that are specific to an instrument are implemented as a "microservice" running on an ESO server and interfaced with ObsPrep via APIs. The allocation of the fibers of a multi-object spectrograph to the objects from a catalogue, or the calculation of the performance of an adaptive optics system based on the position of the reference stars relative to the target are examples of such microservices.

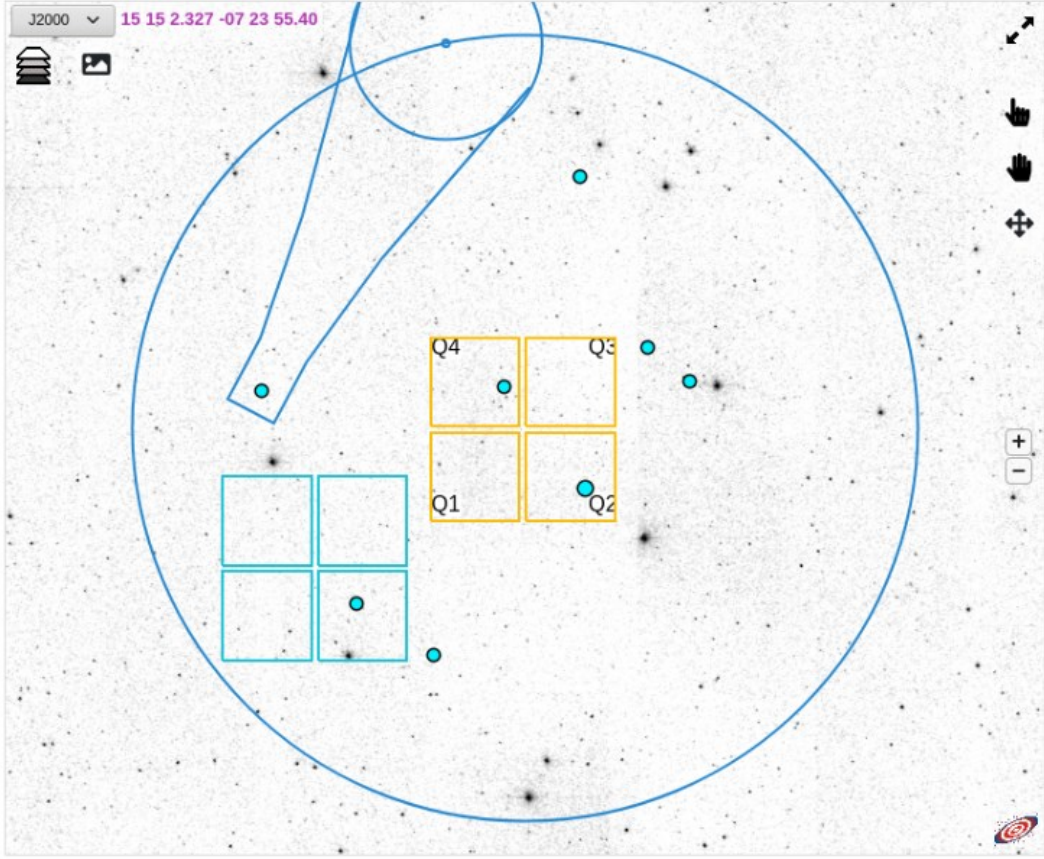

Figure 8. A screenshot of ObsPrep, the tool used for instrument-specific observation preparation within P2 at ESO. Suitable guide stars are plotted over the Digital Sky Survey. The user can position the guide probe on one of them checking that it does not vignette the instrument's field of view (in orange) and select the position of an offset for measurement of the sky (in light blue).

The UPP blurs the line between Phase 1 and Phase 2. Indeed, when a user prepares his/her proposal, if he/she accepts the automatically generated, default, observation in the "*Prepare*" application, that observation already contains all necessary Phase 2 information under the hood, including calibrations. If the default parameters are acceptable, then the user can simply change the status of the observations from "Approved" (by the TAC) to "Ready" (to be scheduled). Automatically generated sequences do not require human review, so these observations will be immediately available for the nightly scheduling process.

In the case that the full automated configuration of the observations needs to be customized, the "*Prepare*" allows modification of the instrument's parameters as well as details of the observing sequence.

The Sequence Editor panel (see Figure 9) displays the input parameters and the resulting sequence of observations in the OB (including the acquisition and on-sky calibrations for facility instruments). Here the user can see exactly what they will get when the observation is executed. Editing basic parameters such as spatial or wavelength dithers will cause the sequence to be automatically regenerated. Again, automatically generated sequences can be moved to "Ready" status directly without human validation. However, once a sequence has been manually edited human validation is required before its status can be set to "Ready."

In addition, each automatically generated observation has associated calibration observations also generated automatically. The associated calibrations themselves have limited user configurability to minimize opportunities for errors. Finally, the Observations may be combined into OR groups and AND groups, which may be nested. An "OR" group allows inclusion of a larger sample of observations to facilitate scheduling. "AND" groups allow the users to request relative timing constraints between the observations in the group (see Figure 10).

Figure 9. A mockup of the Observation view of the GPP/Explore application with a single observation highlighted and showing the Sequence Editor View.

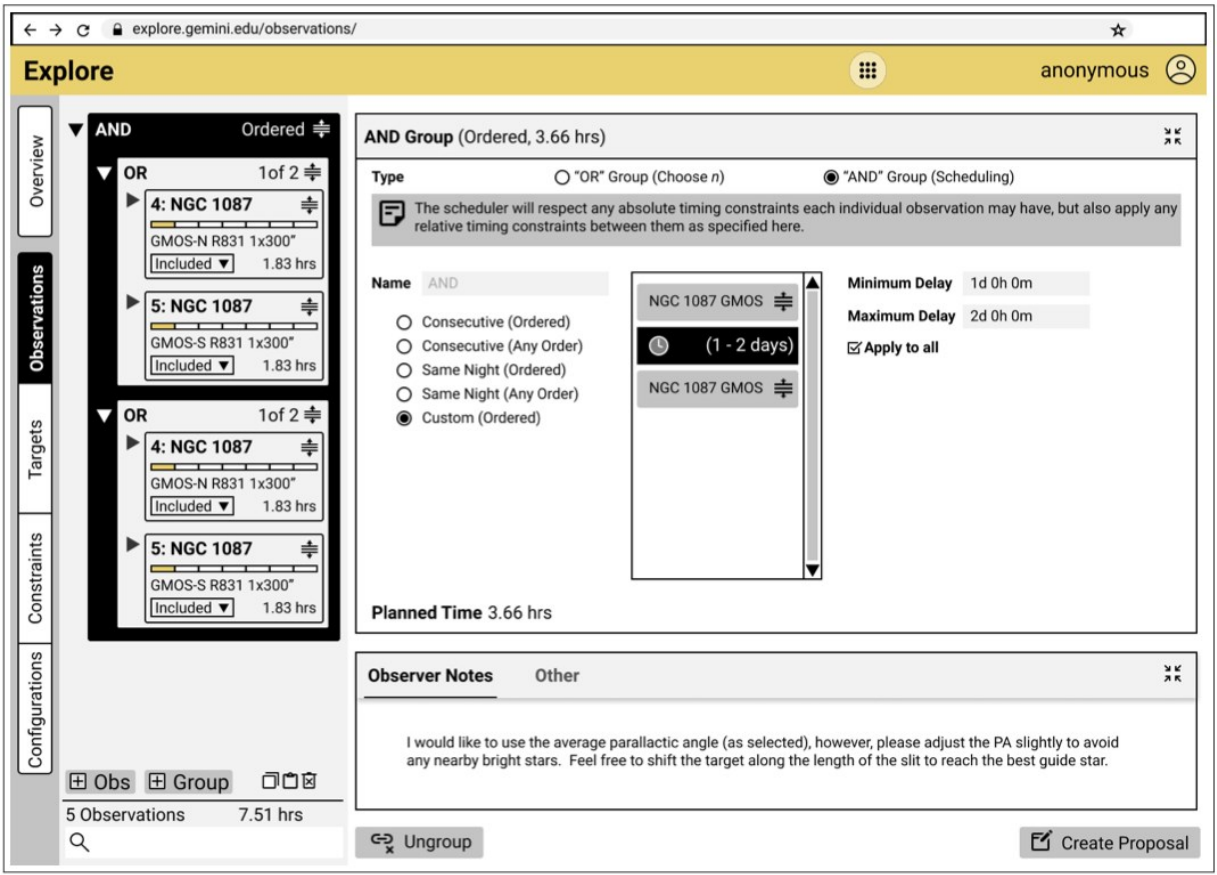

Figure 10. A mockup of an AND group of nested OR groups in the Observation view of the GPP/Explore application.

Additional information like the science targets' description and finding charts can be edited in the "target" view of the Explore application during Phase 2 (see Figure 11). All the features and the Phase 2 process described above, will be adapted to meet GMT and TMT policies.

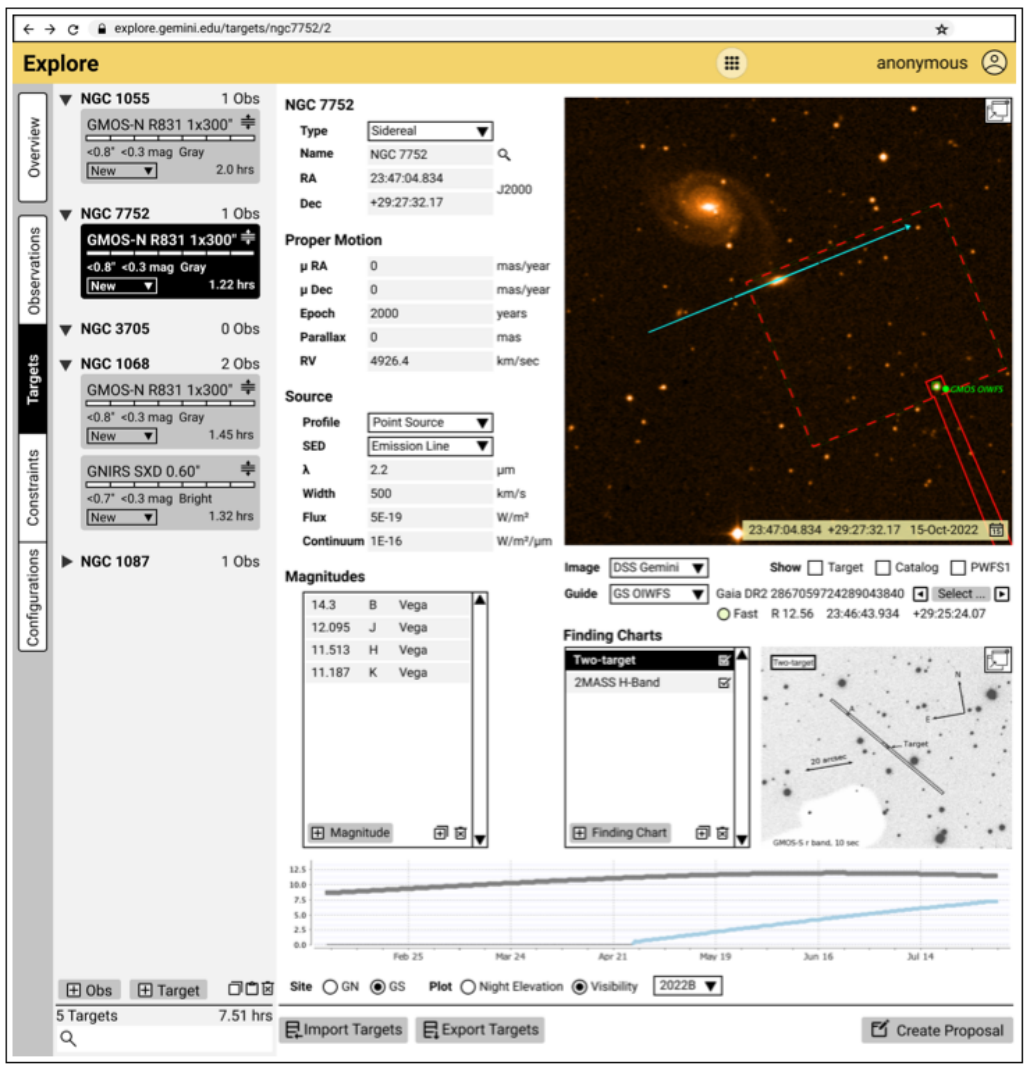

Figure 11. A mockup of the Targets View of the GPP/Explore with an observation selected.

## g. *Short-term scheduling and observation execution*

In Service Mode, (SM) however, the staff observer must filter and rank a large number of OBs belonging to numerous programs on various instruments, taking into account their individual visibilities, constraints (in terms of seeing or turbulence characteristics, moon illumination, humidity, etc., but also time constraints for time-critical observations), and priorities, and pick the most suitable one for execution. The observatory staff performs the observations and evaluates its quality and success. It is possible to alter the planned observations (for example to adjust for the brightness of a flare in a ToO, or to compensate for cloud requiring a longer exposure time than originally planned), but SM observers use this capability sparingly and responsibly.

In Visitor Mode (VM), the detailed scheduling of a night of observation is performed by the visiting astronomer, possibly supported by tools provided by the observatory.

Intermediate modes are also considered, where the scientists requesting the data can interact with the observatory staff and contribute remotely to the observations. For a variety of reasons –not the least of which, for safety—the remote observer can only watch and talk, but not directly act upon the observations; this mode is therefore often called *eavesdropping mode.*

At ESO, in VM, this is done by the visitor populating an "execution sequence" in the P2 tool (see Figure 12 for a screenshot), with OBs that will be executed one after the other. The execution sequence can be updated at any time; thanks to the real-time bidirectional database replication, this update can be done on-site, but also from any other place, enabling remote observations. Furthermore, the execution sequence can be filled and controlled using its APIs, allowing complex strategies to be implemented via program-specific scripts. This enables a *delegated Visitor Mode* (dVM), which is essentially a VM program performed in eavesdropping mode, useful for very short and/or repetitive VM runs.

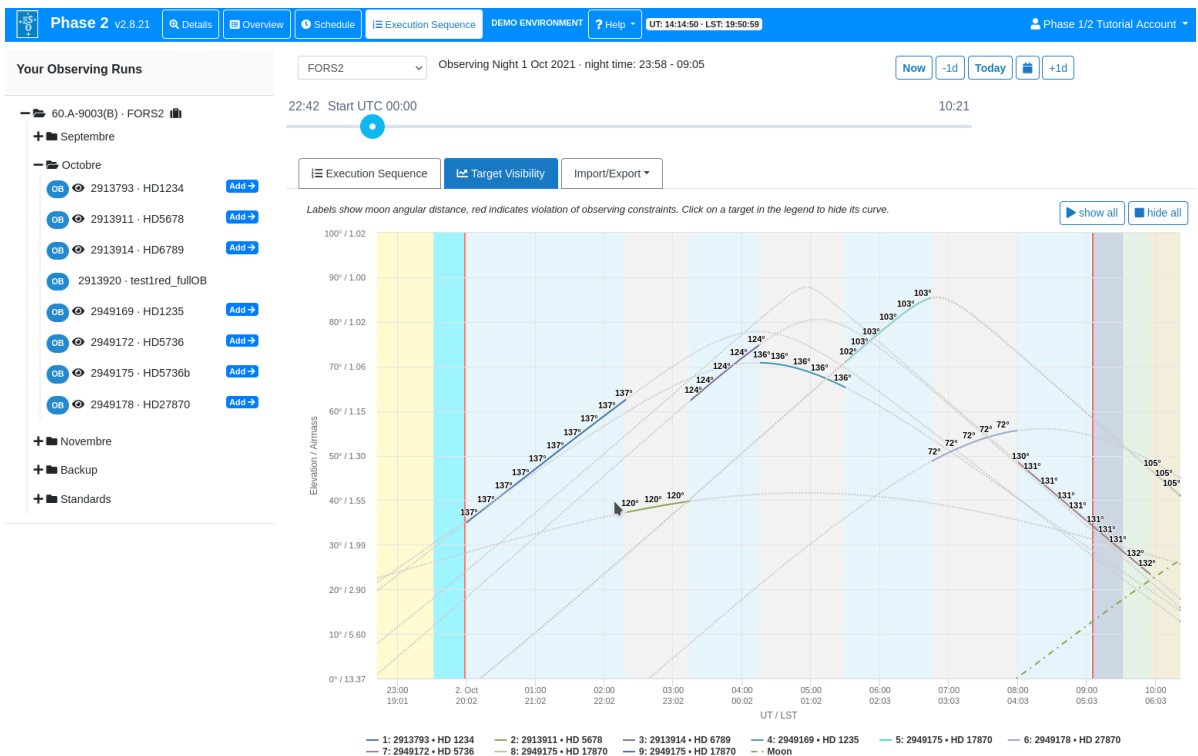

Figure 12. Screenshot of a Visitor Execution Sequence in ESO's P2 tool, showing the airmass of the various OBs waiting to be executed in VM

In SM, the observatory staff filters the OBs that are currently observable (in terms of visibility and observing and time constraints) and rank them. The ranking accounts for priorities, but also for the scarcity of the observation conditions needed (so that an OB requiring, for instance, 0.5 arcsec seeing will have a much higher rank than one satisfied with 1 arcsec seeing) and of the remaining suitable observing time until the end of validity of the OB. Hence, the OB(s) surfacing to the top of the list (higher rank) are the most important and most urgent ones. While at this time only the current observing conditions are considered, ESO is planning to include short-term weather forecasts in the selection. For observations for which this is relevant, collisions with the laser beam from another telescope are also considered, accounting for the rules of engagement and priorities among the various ongoing programs. Finally, once an OB is selected for execution, it is passed to the instrument control system, where a sequencer running on the instrument workstation will execute the various steps described by the templates in the OB. The instrument returns a signal indicating that the OB is completed (or aborted, if appropriate), and sends all the data files produced by the observations back to the DFS. This short-term scheduling is currently performed with the Observing Tool (OT, see Figure 13 for a screenshot), a Java desktop application. Its web-based replacement is being developed, including new ELT requirements, and addressing technology obsolescence. It is scheduled for deployment in 2023.

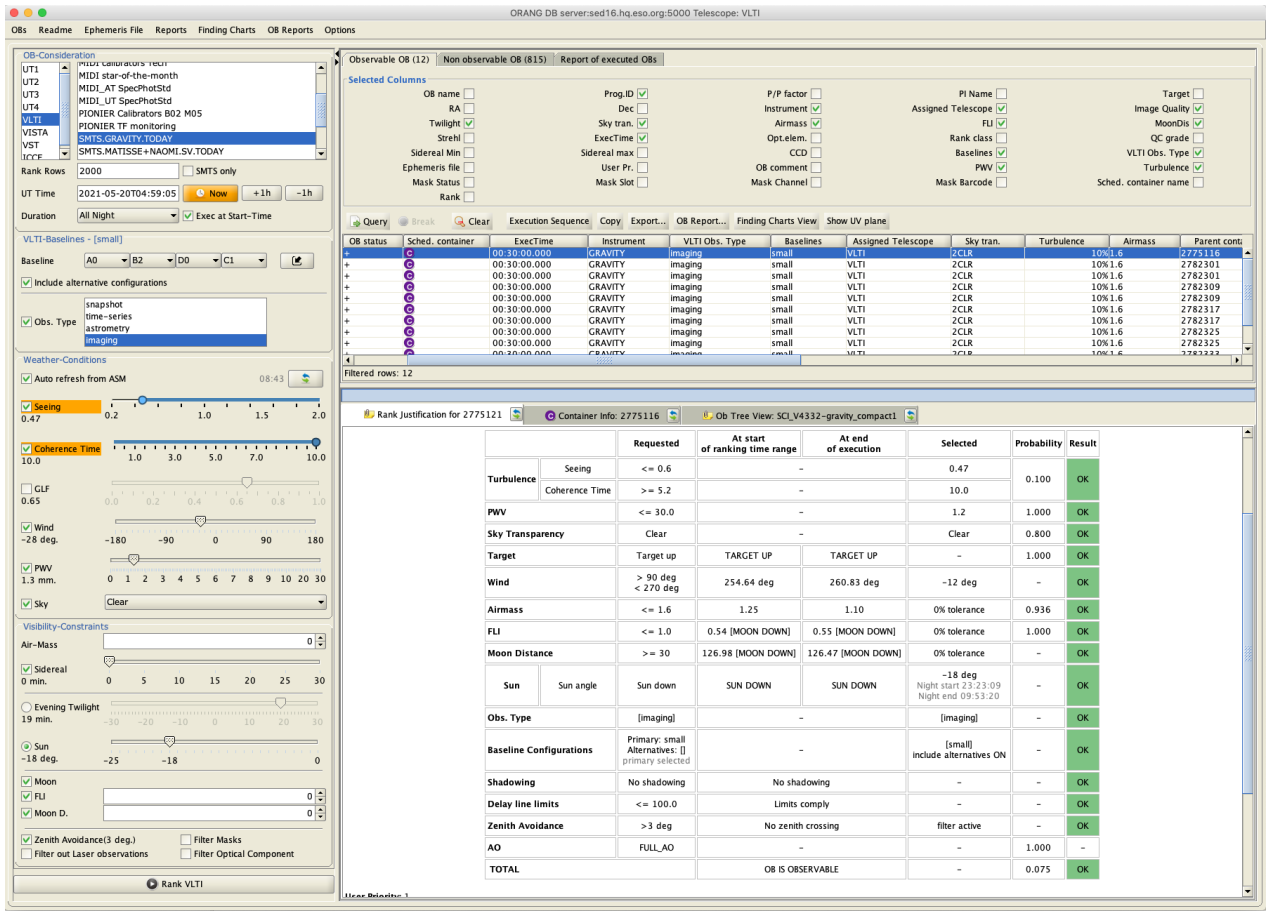

Figure 13.  A screenshot of the current VLT short-term scheduler, OT, showing a list of filtered and ranked observations. The current conditions are listed on the left column. The main window shows the ranking justification for the top observation.

At TMT, the observations will be supported by one Support Astronomer (SA), based at the sea-level control room, and responsible for the instrument operations, the data acquisition, data quality control, reporting, etc, and by two Operation Associates (OA), operating the telescope and AO/LGS, and monitoring the weather conditions from the summit. Together, this team will support the execution of science programs according to the two observing modes offered, SM, and VM.

In SM, the observation selection from a queue of programs is prioritized using an adaptive scheduling engine (see below). In VM, both the visiting astronomer and the SA are located at the sea-level headquarters. In *eavesdropping mode, o*nce an observation is included in the nightly plan the users will receive a notification as well as a second notification if the observation is still in nightly plan no less than 15 mins before the slew time. In that case, the remote users connect with the SA and assists with target identification, acquisition, or even real-time changes in observing strategy. The proportion of time used for each of these two modes will be determined by each TMT partner, based on what its user community sees as an optimal balance between visitor and service modes.

The adaptation scheduling engine, in addition to providing priority ranking based on current and forecast weather conditions, ranking and completions of the programs, also balances the distribution of weather conditions and completed programs among partners, while minimizing telescope slewing time and instrument changes. The adaptive scheduling software will be provided by NSF's NOIRLab and will built upon NSF's NOIRLab/Gemini's 20 years of experience of queue[22]. One of the main requirements of this scheduling software is the automation and adaptation of service mode programs prioritization within minutes as events occur and as observing conditions change during the night. To help with the optimization of telescope schedule, this software will also be able to create observing plans over multiple nights to several months in advance in simulation mode. The SA has the option to either follow the guidelines proposed by the scheduling software or bypass them if they decide that a different program should be executed instead (this decision should be documented as part of the night-report).

At GMT, the entire software framework, including the observation execution software, is highly integrated. NSF's NOIRLab's software will interface with the GMT software mostly at the level of Phase 2, where the NOIRLab-supplied Phase 2 information is converted into GMT-specific execution sequences. A queue scheduler will use Phase 2 information for pending observations, combine it with current and predicted environmental conditions, instrument availability, visibility, etc., and prioritize the pending observations. GMT is experimenting with genetic algorithms for queue optimization. A GMTO Observer will always be available to either advise the VM observing team or to perform the service observations from the queue schedule. We anticipate that VM observers will use the same queue schedule system, but with a switch set to only show the team's program observations and protect proprietary information on other programs.

*h. Observation Log*

The historical observation logs, kept in large notebooks filled by hand, have long been superseded by digital logs filled directly using authoritative information from the instrument and telescope control systems. The detailed information about a given observation is stored directly in the FITS file, in the form of keyword-value pairs. A series of standards defining these keywords are in place, ensuring a good level of interoperability across data from different observatories. The ELTs will of course continue adhering to these standards.

At ESO, the details about the observations are collected by the *Night Log Tool* and stored in a database. These include the OB details, the outcome of the online quality control (QC0, see

Section l below), and comments by the observer. Combining the information from this database with additional queries to the weather station (including measured seeing, and earthquake report if relevant) and to the problem reporting system, custom reports are generated for the VM observer, for the PIs of the programs executed in SM, for the observatory staff... Alternatively, the SM program PIs can monitor the progress of their observations from a dedicated interface accessible via the User Portal, as illustrated in Figure 14.

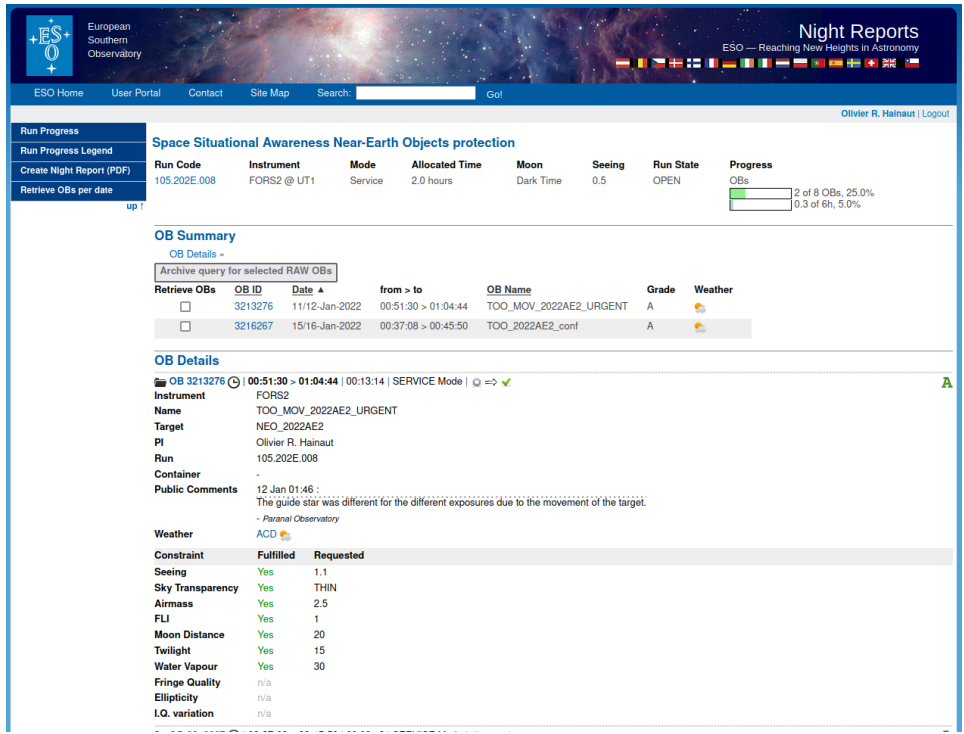

Figure 14. Screenshot of a SM observation report produced by the ESO Night Log Tool.

For TMT, NSF's NOIRLab will provide software that will log all day and night science operations activities into a database system from which will be extracted various operations- or user-oriented reports Figure 15 shows a mockup of a tool, *Chronicle*, (designed by NSF's NOIRLab/Gemini) to extract the night-time report. The information included in the report can be customized according to the recipient of the report (science operations staff, management, science user). For instance, the night view of the Chronicle is designed for use by the night-time science operations team. This section starts off with the usual fields for the night crew identification and other members who have provided additional support, and a brief summary of the night. There is a collapsible Timeline section which includes an interactive elevation plot of the executed observations which builds up as the night progresses and is color-coded by instrument. This provides an overview of all the observations, with vertical bars indicating time lost to weather, faults, laser shutters (due to satellites or airplanes), or spent on engineering, commissioning, or shutdown. Hovering over an observation shows a tooltip with details about the observation and selecting an observation in the plot will highlight the timespan of the observation in the detailed section below. Double-clicking an observation will open it in the Phase 1 & 2 tool "Prepare." Below the timeline is a horizontal bar chart which breaks down the time usage for the night between science, weather, faults, engineering, laser shutters, commissioning, shutdown, and unused (idle time between programs). On laser nights this will also display the time the laser was shuttered for airplanes or satellites. The details section is a mix of lines added by the Night Crew and automatically populated lines with information from various sources. Each line has the time the event started, the "User" who made the entry, and a description of the event. Finally, Operations metrics can be accessed through "Chronicle" to establish performance statistics of the science operations.

← → C  🔒 chronicle.tmt.org/20320124/observing/night  ⭐

**Chronicle**   HST ⚙     ◀ 2032-Jan-23/24 ▶     ▦   SA_W ⊚

**Observing**

▶ Morning  ▶ Afternoon
▼ Night

**Night Crew**
( Gelys T. ⊗ ) ( Marie L. ⊗ ) ( Warren S. ⊗ )

**Additional Support**
( Kyle Koa ⊗ ) ( Rolando R. ⊗ ) ( Jason W. ⊗ )

**Summary**
A night with good seeing, but increasing thickening clouds. Observed a full set of Q-101 observations.

▼ Timeline

☾ 75%

— WFOS (51%)
— IRIS (40%)
— MODHIS (9%)

🔍➖ 🔍➕ ◀ ▶

☐ Fault Loss 0.42 hrs (5.1%)   ☐ Weather Loss 0.92 hrs (11.2%)   ☐ Science 6.81 hrs (83.6%)   Full Night 8.15 hrs

| HST | User | Event | Step | Type | QA |
|---|---|---|---|---|---|
| 19:30 | almanac | Sunset | | | |
| 19:30 | OA | Tuning M1... | | | |
| 19:35 | envmon | IQ=0.72" 0.7um NNE 85° | | | |
| 19:50 | almanac | Nautical 12 degree twilight | | | |
| **32A-Q-232 [38]** | **WFOS** | | IQ < 0.8" CC < 0.1 mag BG 50 WV any | | |
| 19:55 | db | Slew to 32A-Q-232 [38] NGC 1024 | | | |
| 22:01 | fits | 20320123S0028 | 001 | Ⓠ | 🟢 |
| 22:03 | tcs | P=0.92" Q=-0.32" | | | |
| 22:03 | SCMS | IQ=0.59" 0.63um SSW 72° | | | |
| 22:04 | fits | 20320123S0029 | 002 | Ⓠ | 🟢 |
| 22:06 | OA | Centering between the two clumps of emission | | | |
| 22:06 | tcs | P=-0.17" Q=-0.03" | | | |
| 22:07 | fits | 20320124S0030 | 003 | Ⓠ | 🟢 |
| 22:09 | fits | 20320124S0031 | 004 | Ⓠ | 🟢 |
| 22:11 | fits | 20320124S0032 | 005 | Ⓢ | 🟢 |
| 22:11 | db | └Start exposure | | | |
| ☁ 22:21 | db | └Pause exposure: cloud passing overhead | | | |
| 22:26 | db | └Resume exposure | | | |
| 22:31 | db | └End exposure | | | |
| ☁ 22:32 | OA | Start weather loss: What we thought was just a seeing bubble has turned into downright poor seeing that is worse than this program can take. | | | |
| 22:32 | fits | 20320124S0033 | 006 | Ⓢ | ○ |
| ↳ 22:42 | OA | The clouds got thick ~half way through this exposure. (comment associated with file 33) | | | |
| 22:42 | SCMS | Band of thin cirrus passing overhead. Extinction 0.2 mag | | | |
| 22:45 | SCMS | Cirrus getting thicker. Extinction 0.5 mag | | | |
| 22:50 | OA | It doesn't look like the clouds are improving, so calculating a plan for these conditions... | | | |
| 22:53 | fits | 20320124S0034 | 007 | Ⓠ | 🟢 |
| **32A-Q-110 [29]** | **IRIS** | | IQ < 1.0" CC < 0.8 mag BG 50 WV any | | |
| 22:55 | db | Slew to 32A-Q-110 [29] NGC 2048 | | | |
| 23:05 | OA | FR 40936: Start time loss | | | |
| 23:15 | OA | We can't find the guide star. It looks like a problem with the TCS. Calling the AOM for help. | | | |
| 23:30 | OA | We rebooted the TCS and re-slewed and things look much better. Resuming acquisition... | | | |
| 23:31 | fits | 20320124S0035 | 001 | Ⓠ | 🟢 |

**Nightlog** · **Summary** · **QA** · **Time Acct** · **Time Loss** · **All Events**

⌨ **Comment**                                                      ☐ Private
[                                                              ] 📎 Add

☁ Start Weather Loss   ⚙ Start Engineering   ⊘ Start Commissioning   ⊘ Start Shutdown   ⚠ Fault Report

Figure 15. TMT Nighttime Observing log mockup example, based on the GPP/Chronicle.

*i. Laser collision*

Laser observations require extra precautions to protect aircraft and satellites, and even the beams of other telescopes from the laser light. The ESO ELT and GMT will use the VITRO software for aircraft to anticipate and protect aircraft. GMT and TMT will also use a Transponder-Based Aircraft Detection (TBAD) system to detect aircraft passing near their beams. GMT and TMT will coordinate with the Laser Clearinghouse (LCH) to protect against inadvertent illumination of satellites. The information from LCH will be processed by software based on that at existing LGSAO observatories like Keck and NSF's NOIRLab/Gemini (see Figure 16) in order to minimize the overheads involved in shuttering the laser due to passing satellites. Coordination with nearby observatories will adapt the Laser Traffic Control Software currently in use at other observatories, e.g. on the ESO VLT where it is used to avoid collisions between the beams of the nearby telescopes.

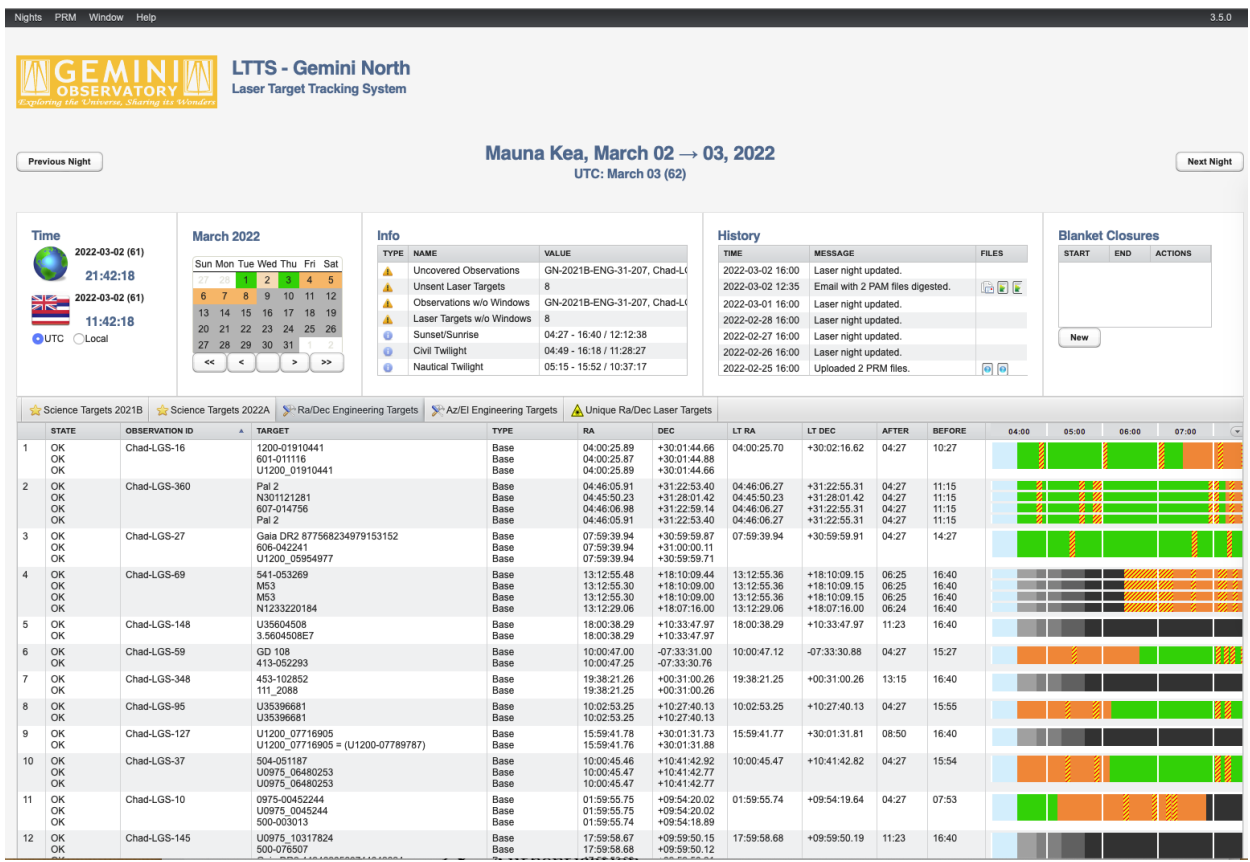

Figure 16. NSF's NOIRLab/Gemini LTTS to automatically send target requests to LCH and automatically process the clearance windows for each target from LCH.
.

*j. Archive and Phase 3*

Once the data are acquired by the instrument, they must be distributed and archived.

At ESO, a local distribution mechanism (the Data Handling System, DHS) takes care of making the data files available to the various systems requiring them, in particular to the observer (for evaluation), to the online data processing pipeline (see below), and to the Data Transfer System, which will transfer the data to the central Science Archive Facility (SAF) at the Garching headquarters in Germany. While the requirement on the bandwidth between the observatory and the archive specifies that 24 h worth of data are transferred and archived within 24 h, most frames are available in the archive within 15 min. In case of urgent observations (e.g., a target of opportunity), a mechanism can boost their priority. The daily volume of data produced by the ELT is expected to be significantly higher than that of the VLT (driven by the requirement to save a significant amount of AO telemetry data), so the data link between the observatory and the headquarters will be upgraded.

The archive holds all the raw data and metadata collected since the early days of the DFS in the late 1990s. The storage system has been updated and upgraded several times and is likely to continue to evolve following the advances of storage technologies and possible new requirements. Originally, the archive was conceived as the data vault in which the observations would be preserved; over time, its role has been expanded to serve as the distribution center for all raw and processed data.

Indeed, the archive also contains processed data, including advanced, science-ready datasets produced either by the community (typically survey and large program data, often tailored to a specific science case) or by ESO (processed using its pipelines to remove instrument signature from the data, and to produce generic science-ready products, see below), via the Phase 3 process[7].

In addition to the traditional web forms where users can specify their queries to the archive, a new interactive interface — the ESO Science Portal (see [23] and Figure 17)— has been developed, from which the user can explore the archive using a physical description of the data among all the instruments and telescope data archived (now also including ALMA), e.g., using wavelength and resolution, rather than filter or grating number. This was made possible thanks to a comprehensive data model standard adhered to for all Phase 3 processed data ingested in the Archive. The Science Portal also shows interactive previews of both images and spectra, allowing the user to evaluate them without downloading the data. The archive can also be accessed using a programmatic interface based on the Virtual Observatory standards.

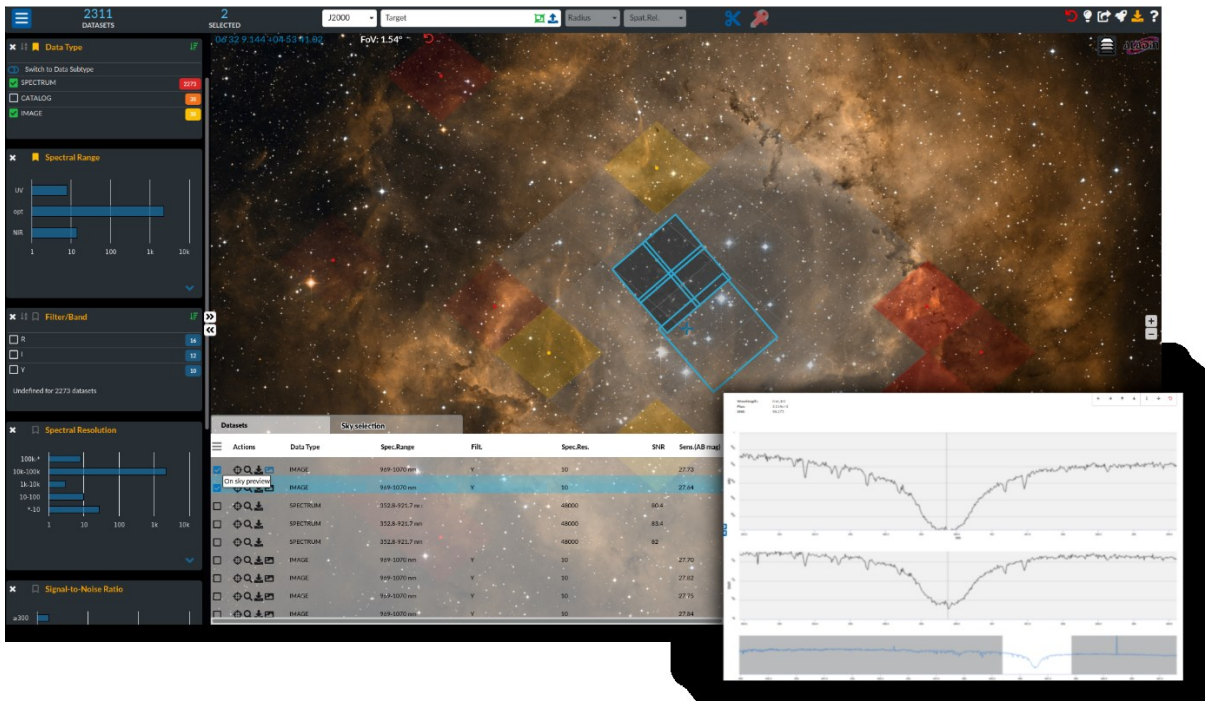

Figure 17. A screenshot of the ESO Science Archive Portal. Assets can be searched based on their location, and on the physical properties of the data. Previews are available for most data types.

GMT and TMT Science data will be transferred in near real-time to the US-ELTP Science Archive at NSF's NOIRLab. This will be the main access point for all science users, including archival researchers. The US-ELTP Science Archive will be an integrated archive for data from both GMT and TMT, and will have a common interface with the other NSF's NOIRLab Data Archives. As for its other Science Data archive, NSF's NOIRLab will implement key International Virtual Observatory Alliance (IVOA) protocols for data discovery and access in the US-ELTP Science Archive.

GMT will retain all raw science data from instruments on the mountain for at least as long as that instrument is in service. This is to facilitate access to historical performance data even if the internet is not available from the mountain. A long-term (50-year) archive will be maintained either at the Base Facility in La Serena or the U.S. Science Operations Center. Redundancy and backup systems will ensure that data are not lost. Newly acquired science data should be available within an hour (goal of 5 minutes) of the end of the observation.

A separate engineering archive will record and curate Observatory subsystem telemetry. This archive will not be copied to NOIRLab as its main purpose is to serve as a diagnostic and trending tool for troubleshooting and maintenance.

TMT uses its Data Management System (DMS) for handling, accessing, and archiving all TMT data over the lifespan of the observatory, which includes both science and engineering data and any external archives established by TMT partners for their own communities.

In its baseline scenario, the DMS contains 2 major archives: The TMT Science and Engineering Archives and the US-ELTP Science Archive. Immediately after their collection, the TMT raw science data will be automatically transferred to the TMT and US-ELTP archives. The calibration files associated with all instrument modes used during the night of observing will be generated the day after and will also be immediately ingested into the TMT and US-ELTP archives.

In addition to the US-ELTP Science Archive, NSF's NOIRLab will develop a science platform, Data Reduction and Analysis Workspace (DRAW) as part of the UPP applications, to serve a large and diverse community of researchers by providing a collection of tools and interfaces uniquely suited for exploring, visualizing, and analyzing data from GMT and TMT. NSF's NOIRLab will build up on its current science platforms (see Figure 18) which have integrated collection of high-level tools and interfaces that support discovery, exploration, visualization, and analysis of astronomical data from Rubin, Gemini and the Middle Scale Observatories in an online environment that is co-located (either physically or virtually) with the Science Data Archive. Similarly, in the UPP/DRAW there will be pre-installed and pre-configured tools and software so that users do not need to manage their own installations. This capability will facilitate research inclusion because scientists at under-resourced institutions often have limited computing capability at their home institutions. The specific tools to be provided for working with data produced by GMT and TMT will be defined in collaboration with scientists who plan to use these telescopes and will take advantage to the greatest extent possible of the resources already provided by NSF's NOIRLab. A basic set of capabilities will likely include an online notebook environment (e.g., Jupyter or another newer technology), access to large astronomical catalogs including imaging and spectroscopic data sets through image cutout services, cross-matching, and personal allocations of file storage (eg. Astro Data Lab provides 1TB of virtual disk storage per user), database storage (eg. 250 GB of personal MyBD database storage in Astro Data Lab), persistent repository for published data and computing for individual users.

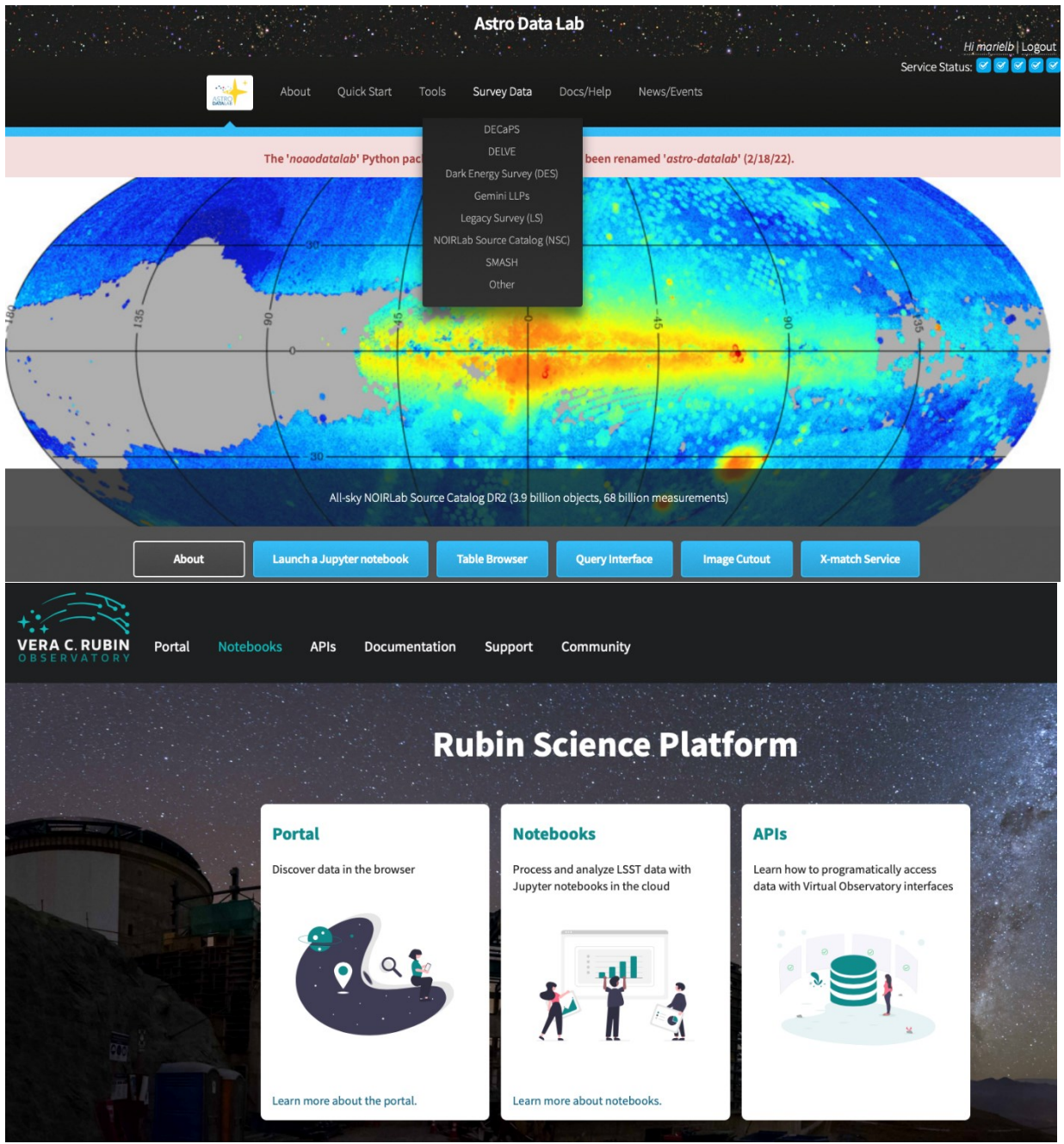

Figure 18. Top: NSF's NOIRLab Astro Data Lab. Bottom: NSF's NOIRLab Rubin Science Platform. Both are the current precursors of the UPP/DRAW.

*k.   Data processing pipelines and infrastructure*

The complexity of the raw data produced by recent and upcoming instruments is continuously increasing. In many cases, gone are the days when a quick look at the frames with a basic exploration tool could reveal whether the observation was successful or not, and when a home-made collection of scripts could process the data and to allow an astronomer to start analyzing them. Data processing pipelines are now a critical component of the instrument.

At ESO, the observatory commits to providing, for every instrument, a pipeline that removes the instrument signature from the data. This includes converting the instrumental units (position in pixels, intensities in ADUs) into physical ones (pixels in sky coordinates or wavelengths, and fluxes in physical units or via a photometric zero point). This also includes re-shaping the data from whichever format they come into (which can be extremely complex for instance in the case of integral field spectrographs or multi-object spectrographs) into 1D, 2D or 3D frames. At the core of the pipelines lies the Common Pipeline Library (CPL), a collection of C functions performing basic tasks, and the High-level Data Reduction Library (HDRL), another series of functions to perform advanced tasks that are common to several pipelines. The pipeline for one instrument is constituted by a series of recipes, which can be called independently (e.g., to create a master flatfield from a series of raw flats), or together via a workflow that connects them in the right sequence and selects the right input for each recipe. Currently, the pipelines are operated via various infrastructures that were developed and evolved to support the various use cases (from the telescope to the final user's computer). In preparation for the ELT, ESO is completely overhauling the infrastructure that calls the pipelines. The new infrastructure will be deployed at the telescope for quick look and quality control, at the headquarters for advanced quality control and production of data products, and on the user's system for final data processing. Also, the CPL is being packaged within a layer that will make it directly accessible from Python. This will allow users (both astronomers and developers) to directly call CPL functions and pipeline recipes from a Python script (with fully pythonic interfaces). The goal is to benefit from the power of the highly optimized CPL recipes in the rich and flexible Python environment. This will also allow developers to prototype new recipes in Python.

GMT and TMT are planning to provide seven first-generation instruments for the two telescopes combined, including both imagers and spectrometers (see "Instrument Program of the Extremely Large Class Telescopes" in this issue). Each instrument will be delivered to GMT/TMT with its set of operational documentation (user manual, calibration plans, data-reduction manual) and data-reduction pipelines. NOIRLab is currently working with GMT/TMT to agree to common standards for data formats, metadata, and environment. The US-ELTP goals is to provide for each instrument a DRP framework with three modes for each GMT/TMT instrument: 1) a quick-look mode designed for rapid turnaround and assessing data quality during observing; 2) an automated mode to run for Standard Data Reduction (SDR), in which little or no user interaction is required; and 3) an interactive mode in which users can interact with and modify reduction parameters. The observatories (including their instrument teams) and NOIRLab will work together to define SDR-supported observing modes, SDR-required calibrations, and the procedures and data products for SDR. DRP recipes are expected to be continuously developed and improved. The evolution and improvement of the pipeline modules involved in the production of science-ready data will be the joint responsibility of GMT/TMT staff, GMT/TMT instrument team members, and NOIRLab staff involved in the development of data reduction software. The science user community also will be able to provide feedback on the DRPs. Pipeline and recipe code will be version-controlled, and changes will only be incorporated after review by all the relevant parties.

*l.   Quality control and advanced data products*

Traditionally, the observer was controlling the quality of the data they just acquired, adjusting their exposure time and strategy accordingly. While this is still the case in VM, the quasi-industrial aspect of SM implies a more formal approach to quality control (QC). Various levels of QC can be considered. QC0, in quasi-real-time at the telescope, as soon as the data are acquired, a first set of tests are performed, and the sequence of observations can be affected (e.g., repeating an exposure that was not up to specifications). QC1 can take place during the day and may affect the following night. Finally, QC2 is a long-term, slow-paced study that will only affect the evolution of the instrument.

At ESO, currently QC0 systematically compares the observing conditions with the specifications provided in the OB (e.g., in terms of seeing) and in some cases performs measurements on the data themselves to grade the observations (A: meet the requirements; B: almost meets the requirements and does not need to be repeated; C: observation failed, must be repeated). QC1 focuses on the stability of the instrument within its specified performance, evaluated from the calibration frames (including parameters such as the readout noise of the detector, the resolution of the spectrograph, the photometric zero-point of the camera...). Discrepancies trigger an alert that is investigated at the observatory, with consequences such as the replacement of a calibration lamp or the adjustment of an optical element. A by-product of this QC1 step is the production of quality-certified master calibrations (flatfields, response curves...) that are stored in the archive, from which they can be associated with the scientific data they are relevant to.

In preparation for the ELT, ESO is now increasing the scope of its QC and developing a new infrastructure that will support each level of the QC process and interface with the new data processing infrastructure described above to run the pipeline. For QC0 at the telescope, the system will also interface with Scuba, the real-time QC0 platform developed at the observatory. QC1 is now taking place at the Santiago office, taking advantage of the availability of the instrument and operation specialists. QC2, at the Germany headquarters, will focus on the science quality control, improvements of the pipelines, and creation of data products. With the development of science-grade pipelines, ESO has embarked on the systematic processing of the data for more and more instruments, generating science-grade data products that are made available in the archive and highly appreciated by the users[6].

At the TMT, once the SM OB has been executed, the support astronomer will provide a preliminary assessment (QC0) of the OB quality by comparing the ambient weather conditions with those requested by the PI. This real time quality control will be complemented by measurements made directly on the image (FWHM, Strehl, Signal-to-Noise Ratio) and comparing them with requirements from the science proposal. As for ESO, this QC0 aims at providing triage of the executed OB between those executed successfully and those which have failed execution (due for instance to the degradation of weather conditions, or instrumental problems). The day after, Observatory staff will be able to provide complementary quality control as needed, like in the few cases when the QC0 assessment is flagged as uncertain.

Once the calibration files associated with the science observations have been generated, they are transferred to the TMT and US-ELTP archive. This step will trigger automatic reduction of the science data using the data-reduction pipelines maintained by NOIRLab. During this process, a final check of image quality (QC1) will be provided using quality control tools applied to both the calibration and science data. The result of the QC1 process applied to calibration data will be fed-back to the observatory and will be used by the science operations staff to certify the calibration and monitor the status of the science instruments. The result of the QC1 applied to the science data will provide PIs with a final verification of the image quality obtained.

At the GMT, the observer will use quick-look reductions to compare the quality of the data to what was requested by the PI. This QC0 can include image quality, but also S/N and other parameters. Quick-look reductions may not use the most optimal calibrations and may skip steps in the reduction process in order to produce a result rapidly. Subsequent data processing using the proper calibrations and all reduction steps will provide a better QC1 assessment and may be carried out the next day. Performance parameters derived from calibration files will be automatically processed separately and compared to expectations, to indicate a potential problem. Since higher level data products will mostly be produced for the US-ELTP archive by automated processes, each will have its own automated QC assessment. Manual (human) QC assessment is time-consuming and expensive and will be used only when necessary.

The US-ELTP will carry out "Standard Data Reduction" (SDR) for science observations from GMT and TMT using Data Reduction Pipelines operated at NOIRLab. The primary goal of SDR is to routinely generate reduced, calibrated data products that can be used by investigators for scientific analysis, and to archive those data products for future use by archival researchers. The observatories (including their instrument teams) and NOIRLab will work together to define SDR-supported observing modes, SDR-required calibrations, and the procedures and data products for SDR.

Each TMT and GMT instrument will be responsible for defining the various data products that will be served through the archive. This will include raw data, potentially including individual reads of a detector (TMT's IRIS). It will also include data that have undergone standard data processing to remove instrument, detector, and telescope artifacts, such as bias correction, flat-fielding, etc. Some instruments, such as IRIS and GMT's GMTIFS, will produce 3-D data cubes (x, y, wavelength).

Higher level data products may include extracted spectra (including individual orders from a cross-dispersed echelle, or individual spectra from a multi-object spectrograph). Combined spectra (from multiple orders) or mosaicked images may be even higher-order data products.

NOIRLab will also accept contributed data sets; data that have been processed outside the UPP environment, but that provide added value, such as combined data from multiple epochs, or data requiring special processing such as precision radial velocities or high-contrast ("extreme") adaptive-optics imaging.

*m.  Closing the loop: publications and other metrics*

The final and fundamentally crucial step of the end-to-end process is the publication of new results in scientific papers. This will foster new questions that will, in turn, trigger new proposals for observing time, closing the loop of the overall process in Figure 1. Therefore, the number and impact of publications produced by a proposal, by an instrument, by an observatory are important and interesting metrics to evaluate their contribution to scientific progress, or more prosaically to measure the return on investment. The field of bibliometrics developed a series of analysis and statistical tools to quantify these estimates. Other aspects of the end-to-end operation process also need to be evaluated and quantified. Some can be measured from the various operation logs, such as "open shutter" duty cycle to indicate what fraction of nights assigned for science operations are spent collecting photons from the astronomical targets. For others, users must be polled, for instance to evaluate customer satisfaction, and to provide specific suggestions for improvement. Finally, in each of the three ELTs, the users will also be represented in a user's committee. Each organization also collates statistics and metrics into reports to their advisory and governing bodies.

The ESO Library and Information Centre maintains TelBib, the telescope bibliography database[24], listing the refereed papers that were produced using partly or exclusively data from ESO, obtained either through dedicated observing proposals, or using the science archive. The selection of the articles and the curation of their attribution to one or another facility is a complex process, described in detail in [25]. This trove of data is used to put ESO's telescope in the global context of other ground- and space-based observatories and also to evaluate the impact of its various instruments, as illustrated in Figure 19.

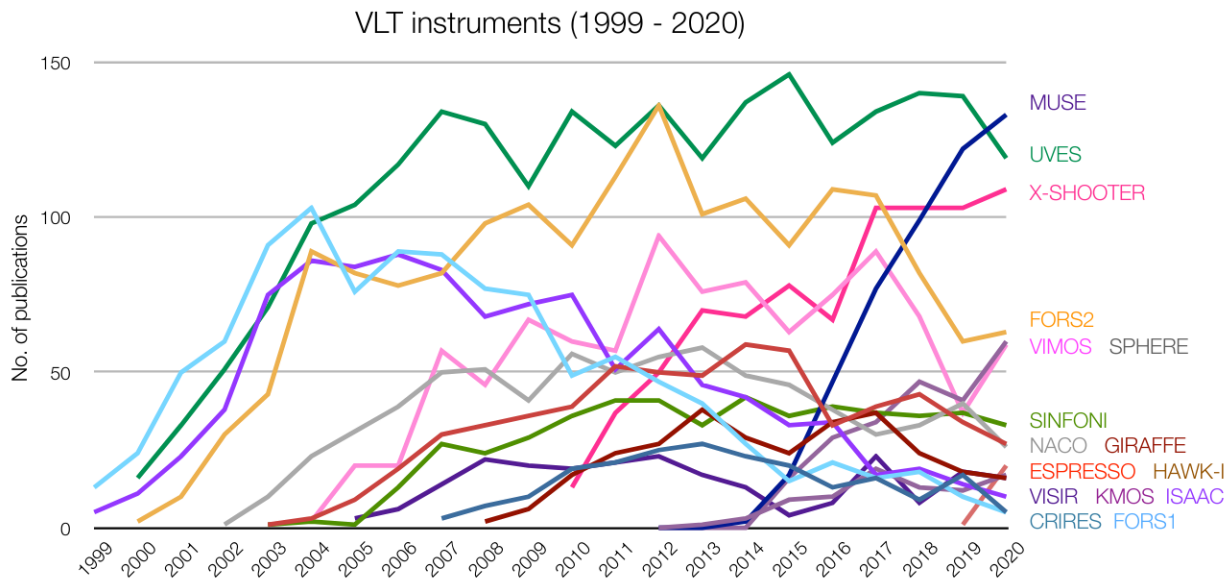

Figure 19. Refereed publications using data from VLT instruments, using TelBib, from [25].

NSF's NOIRLab Library Services (NLS) track refereed publications by observers using NSF's NOIRLab facilities or data products as well as refereed and non-refereed publications by NOIRLab staff. NLS maintain public libraries in ADS: the NOIRLab ADS bibliographic group and the NOIRLab Staff Publications public library. NLS also compile NFS's NOIRLab Publication metrics for reports (see https://noirlab.edu/science/library/publications/metrics). In addition, each NSF's NOIRLab facility like NSF's NOIRLab/Gemini tracks how telescope time is used each night at each site, also monitors the completion rates for observing programs and publish these Science Operations Statistics on the observatory website[26]. NSF's NOIRLab will track the publication of science using GMT and/or TMT data. Having a single team track metrics for both observatories will ensure consistency at least between those telescopes. Such metrics will also allow an analysis of the scientific productivity of the different instruments on a specific telescope. They may be used to help justify retiring an old instrument to provide space for a newer, more capable instrument, or to propose an upgrade to an existing instrument.

In terms of user satisfaction, ESO distributes to users and analyses a series of questionnaires and surveys, ranging from the "end of mission report" filled by visiting astronomers after their observing run, to a yearly "user satisfaction survey" sent by the User Support Department to all users. Ultimately, users are also represented by the User's Committee, one of ESO's governing bodies. Numerous other metrics are collected and analyzed at all stages of the end-to-end process. ESO has set up a platform to collect, display, and analyze these data, the Dashboard for Operational Metrics at ESO (DOME).

In terms of GMT/TMT user satisfaction, NSF's NOIRLab will adapt a similar approach than for NSF's NOIRLab/Gemini. Currently NSF's NOIRLab/Gemini maintains a direct dialog with its users by sending out routine Short Surveys (2–3 questions) at every critical phase of user proposals/programs (Phase 1, 2, end of semester and Phase 3). The effort has several objectives: 1) monitor the usefulness and usability of the Observatory software tools and documentation; 2) determine how well the observations went, and 3) assess how satisfied the Principal Investigators (PIs) are with the data and how much their expectations were met. Another objective is to identify actionable items that can improve any part of the observing process. As the name Short Surveys indicates, the surveys are designed to be short; they should take only a few minutes to complete. Still, for users who want to have more lengthy communications with Observatory staff, the surveys always include one open question offering a text box that has no length limit. Figure 20 shows the compilation of the responses receive between semester 2016B and 2020B (total response rate of 35%) on the challenges that the users are facing that may prevent publication of their scientific data.

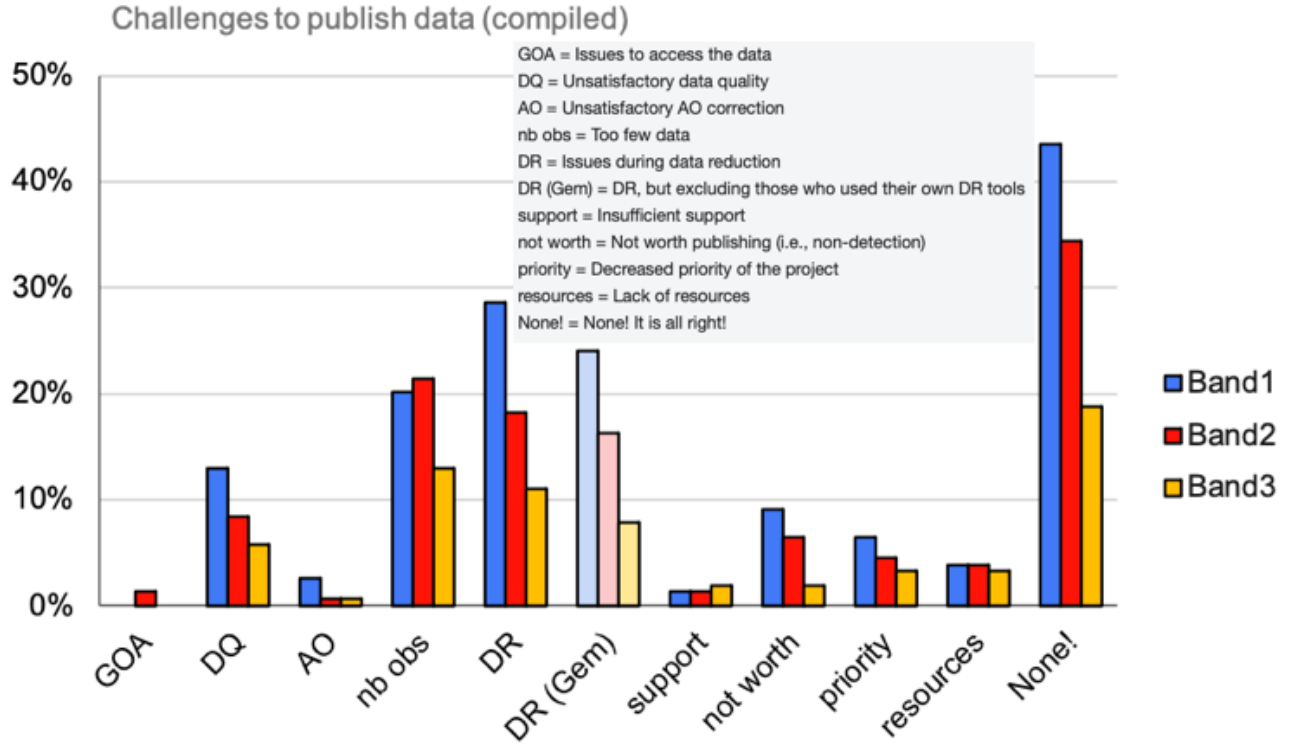

Figure 20. Compilation of the responses receive between semester 2016B and 2020B from NSF's NOIRLab/Gemini on publishing their data.

## 3. Summary and conclusions

While the concept of end-to-end operations is ancient, its implementation as a series of interconnected subprocesses and supporting tools has been key to the success of observations at the major observatories for over 20 years. The development over the years of service and queue observing has improved observing efficiency. In parallel, observatories have been continuously reviewing their operation processes to maximize the scientific production of their equipment, and increasing and improving the level of service provided to their users, reaching the level that was originally typical only for space-based observatories.

Implementing good maintenance strategies using modern tools (see the accompanying JATIS article on maintenance [27]) has helped decrease time lost to instrument and telescope faults, at the same time helping to maintain a high level of performance. The inexorable progress of software technologies has allowed the development of more integrated software systems, as well as the development of more user-friendly interfaces and sophisticated dataflow systems.

All three observatories have evolved towards similar solutions. New requirements emerging from the ELTs and their instrumentation (see [28] for a description of the instrument programs of the three ELTs), from new technical capabilities, new observation strategies, and to cope with software obsolescence, further evolution is needed. At ESO, an ambitious overhaul of the dataflow system is taking place to support its upcoming ELT and integrate its operations with the existing telescopes. The US ELT Program also builds its end-to-end operation process on the strong legacy and experience accumulated during previous and ongoing projects.

This paper highlights the complexity of the overall end-to-end operation process for large observatories serving broad and diverse communities. The differences between the approaches deployed by the ELTs illustrate the history and legacy of various communities and organizations, but the striking similarities suggest an evolutionary convergence, based on experiments, natural selection, and of course the healthy dose of cross-inspiration to be expected among our organizations.

*Acknowledgments*

The European Southern Observatory (ESO) designs, builds and operates world-class observatories on the ground and promotes international collaboration in astronomy. Established as an intergovernmental organization in 1962, ESO is currently supported by 16 Member States (Austria, Belgium, the Czech Republic, Denmark, France, Finland, Germany, Ireland, Italy, the Netherlands, Poland, Portugal, Spain, Sweden, Switzerland and the United Kingdom), along with the host state of Chile and with Australia as a Strategic Partner. ESO operates three observing sites: at Paranal, ESO operates the Very Large Telescope and its Very Large Telescope Interferometer, as well as two survey telescopes, VISTA and the VLT Survey Telescope. Also, at Paranal, ESO will host and operate the Cherenkov Telescope Array South. Together with international partners, ESO operates two millimeter and submillimeter facilities, APEX and ALMA on Chajnantor. At Cerro Armazones, ESO is building its Extremely Large Telescope.

The US Extremely Large Telescope Program (US-ELTP) is a joint effort of three organizations: TIO, GMTO, and NOIRLab. NSF's National Optical-Infrared Astronomy Research Laboratory (NOIRLab) is the national center for ground-based night-time astronomy in the United States and is operated by the Association of Universities for Research in Astronomy (AURA) under a cooperative agreement with the National Science Foundation Division of Astronomical Sciences.

The TMT Project gratefully acknowledges the support of the TMT collaborating institutions. They are the Association of Canadian Universities for Research in Astronomy (ACURA), the California Institute of Technology, the University of California, the National Astronomical Observatory of Japan, the National Astronomical Observatories of China and their consortium partners, and the Department of Science and Technology of India and their supported institutes. This work was supported as well by the Gordon and Betty Moore Foundation, the Canada Foundation for Innovation, the Ontario Ministry of Research and Innovation, the National Research Council of Canada, the Natural Sciences and Engineering Research Council of Canada, the British Columbia Knowledge Development Fund, the Association of Universities for Research in Astronomy (AURA) and the U.S. National Science Foundation.

*Author biographies*

**Olivier Hainaut** is an astronomer at ESO. He has been responsible for operations of the NTT on La Silla, then of the La Silla observatory, then of the VLT. He is now the "end-to-end operations" scientist for ESO. He received his PhD in physics at the University of Liege, Belgium, in 1994.

**Marie Lemoine-Busserolle** is the System Scientist of the US ELT Program at NOIRLab. She received her PhD in Astrophysics at the "Université de Toulouse," France, in 2003.

**Christophe Dumas** is the Observatory Scientist and Head of Operations for the Thirty-Meter-Telescope. He received his PhD in Astrophysics at the "Université de Paris," France, in 1997.

**Robert Goodrich** is the Observatory Scientist for the Giant Magellan Telescope. He received his PhD in Astronomy and Astrophysics from the University of California at Santa Cruz in 1988. He has worked at Keck, Palomar, McDonald, and Lick Observatories as well as STScI.

**Bryan Miller** is an astronomer and Lead Scientist for Science Operations Development for Gemini Observatory at NOIRLab. He received his PhD in astrophysics from the University of Washington in 1994.

-o-